\documentclass[pdflatex,sn-mathphys-num]{sn-jnl}

\usepackage{graphicx}%
\usepackage{multirow}%
\usepackage{amsmath,amssymb,amsfonts}%
\usepackage{amsthm}%
\usepackage{mathrsfs}%
\usepackage[title]{appendix}%
\usepackage{xcolor}%
\usepackage{textcomp}%
\usepackage{manyfoot}%
\usepackage{booktabs}%
\usepackage{algorithm}%
\usepackage{algorithmicx}%
\usepackage{algpseudocode}%
\usepackage{listings}%
\usepackage{xspace}%
\usepackage{subcaption}
\usepackage[utf8]{inputenc}
\usepackage[T1]{fontenc}
\usepackage{anyfontsize}
\usepackage[normalem]{ulem}  
\usepackage{CJKutf8}
\usepackage{longtable} 
\usepackage{siunitx} 

\usepackage{bm}
\usepackage{longtable}
\usepackage{booktabs}
\usepackage{siunitx}
\usepackage{makecell}
\newcommand{\arcsec}{^{\prime\prime}}
\newcommand{\degree}{^\circ}
\newcommand{\rsun}{R_\odot}

\begin{document}

\begin{CJK*}{UTF8}{gbsn}   

\title{Imaging spectroscopy reveals spike-like repeating radio burst pairs in the solar corona}

\author*[1,2,3,4]{\fnm{Suli} \sur{Ma}(马素丽)}\email{masuli@nssc.ac.cn}
\author*[2]{\fnm{Eduard P.} \sur{Kontar}}\email{eduard.kontar@glasgow.ac.uk}
\author[2]{\fnm{Daniel L.} \sur{Clarkson}}
\author[3,1,4]{\fnm{Huadong} \sur{Chen}（陈华东）}
\author[1,3,4]{\fnm{Yihua} \sur{Yan}（颜毅华）}

\affil*[1]{\orgdiv{State Key Laboratory of Solar Activity and Space Weather}, \orgname{National Space Science Center}, \orgname{Chinese Academy of Sciences}, \orgaddress{\city{Beijing}, \postcode{100012}, \country{China}}}
\affil[2]{\orgdiv{School of Physics \& Astronomy}, \orgname{University of Glasgow}, \orgaddress{\city{Glasgow}, \postcode{G12 8QQ}, \country{UK}}}
\affil[3]{\orgdiv{National Astronomical Observatories},\orgname{Chinese Academy of Sciences}, \orgaddress{\city{Beijing}, \postcode{100101}, \country{China}}}
\affil[4]{\orgdiv{School of Astronomy and Space Science},\orgname{University of Chinese Academy of Sciences}, \orgaddress{\city{Beijing}, \postcode{100049}, \country{China}}}

\footnotetext{This manuscript has been accepted for publication in \textit{Nature Communications}. The final version of record will be available at \url{https://doi.org/10.1038/s41467-026-74137-2}.}

\abstract{Solar radio bursts exhibit complex fine structures that reveal intricate coronal plasma dynamics. Here, we report detection of spike-like repeating burst pairs, characterized by two short-lived (0.1-2 s), narrowband components separated by about 4 s at frequencies 30-50 MHz. Using high-resolution dynamic spectra and spectroscopic imaging, we analyzed 613 burst pairs, measuring their durations, bandwidths, drift rates, flux densities, and spatial characteristics. Imaging links sources to an active region, with earlier components spatially concentrated above the region while delayed components are displaced and exhibit reduced drift rates. Radio-wave propagation simulations support the delayed bursts as turbulent echoes of harmonic emission in anisotropic coronal plasma. The location of the burst sources high in the corona suggests ongoing magnetic reconnection and electron acceleration well above typical flare heights. Our findings offer new insights into coronal turbulence effects while advancing diagnostics of coronal plasma and the elusive nature of solar radio echoes from ground-based transmitters.}

\maketitle
\end{CJK*}

\section*{Introduction}
The solar atmosphere is a turbulent and magnetized environment, with the release of magnetic energy readily manifesting as emission across the electromagnetic spectrum.
Solar radio emission dominates the radio sky, with the brightest solar radio bursts generated via the plasma emission process \cite[e.g.][]{Melrose1980}. 
This emission occurs close to the local plasma frequency $f_\mathrm{pe}$ and/or its harmonic $2f_\mathrm{pe}$. Accelerated non-thermal electrons traveling outward from the Sun generate plasma emission at the local plasma frequency, $f_{pe}\propto \sqrt{n_e}$, which decreases with the ambient electron density $n_e$. 
Consequently, solar radio bursts typically drift from high to low frequencies as the electrons propagate through the corona. This frequency drift, observed in dynamic spectra, provides a direct diagnostic of the changing plasma environment along the burst's path. Among the various broadband burst types, sub-second structures may appear embedded within these bursts or occur as isolated features in time and frequency. 

Radio flux variations along the envelope of type III bursts form so-called type IIIb emission, with the individual fine structures referred to as striae \cite{delanoe1972}.
Striae are characterized by durations of less than 1~s, individual bandwidths in the tens of kilohertz range, and drift rates from zero up to a few hundred kilohertz per second at decameter frequencies \cite{Baselyan1974, delanoe1975, Kontar2017, Sharykin2018}. 
When these fine structures are isolated from any broadband burst, they appear as narrowband, short-lived increases in flux known as spikes. These spikes are generally pseudo-randomly distributed across dynamic spectra and have been associated with fragmented energy release in solar flares \cite{Benz1985}.
Spike radio bursts at and above decimetric frequencies \cite{Staehli1986, Benz1992, Huang2005, Dabrowski2005, Rozhansky2008, Huang2022} are linked to maser emission (or plasma emission \cite{Tan2013}) and exhibit millisecond durations with fast drift rates up to hundreds of megahertz per second. 
In contrast, decameter spikes \cite{Barrow1994, Melnik2014, Shevchuk2016, Clarkson2023} have longer durations (up to about 1 s), lower drift rates (from zero to tens of kilohertz per second), and bandwidths similar to those of striae. 
The similarity between low-frequency spikes and individual striae suggests a common formation mechanism of plasma emission \cite{Clarkson2021},  with the primary driver being weaker electron beams than those responsible for type IIIb bursts \cite{Melnik2014}.
Two parallel drifting narrow-band stripes (drift-pair bursts) temporally separated by about 2 seconds with drift rates of approximately $1-8$~MHz~s$^{-1}$, \cite{Roberts1958, Ellis1969, delanoe1971, Abranin1977, Melnik2005, Kuznetsov2019} are also observed.
The repeated burst is considered a turbulent echo of the first formed via radio-wave reflection of the fundamental emission lower in the corona \cite{Roberts1958, Kuznetsov2020}.

Radio emission propagating through turbulence experiences scattering effects. These propagation effects are particularly important for radio emissions at the fundamental or harmonic \citep{Kontar2023}, 
since the refractive index $n_\mathrm{ref} = (1 - f_\mathrm{pe}^2 / f^2)^{1/2}$ is $ \ll1$, where $f$ is the radio emission frequency.
Observationally, radio bursts exhibit significant scattering effects, including broadening in time \cite{Krupar2018} and size \cite{Kontar2017}, delayed arrival times \cite{Chen2023}, and a spatial displacement from the emitting source location \cite{Kontar2017, Kuznetsov2019, Clarkson2021, Clarkson2023}. 
The density fluctuations are thought to be field-aligned (elongated along the magnetic field direction) so that the scattering process is anisotropic.
The level of anisotropy $\alpha=q_\parallel/q_\perp$ (where $q$ is the wavenumber of the density fluctuations axially aligned to the magnetic field) is typically between $0.25-0.4$ \cite{Kontar2023} where $\alpha=1$ is isotropic and has been shown to explain all observable characteristics of radio burst emission simultaneously \cite{Kontar2019}. 
The comparison of the radio and in-situ measurements \citep{2025ApJ...991L..57K} suggests that ion-scale turbulence responsible for radio-wave scattering is consistent with kinetic Alfven waves or kinetic Alfven wave structures \citep{2019ApJ...879...82P}.
Anisotropic scattering leads to a focused channeling of the radio emission along the magnetic field, which can produce directive emission where the apparent source drifts in the direction of the magnetic field \cite{Clarkson2021, Clarkson2023} and can also give rise to the aforementioned radio echo in drift-pair bursts. Stronger anisotropy ($\alpha\lesssim0.2)$ leads to shorter burst duration such that the echo can form an isolated structure or appear within the tail of the primary component if the anisotropy level is weaker \cite{Kuznetsov2020}.
Intriguingly, reflection of Earth radar signals by the Sun is an unsolved puzzle after over 50 years of active research. Unlike planetary reflections, the Sun proves to be a challenge for the radar experiments; the echo at 50 MHz is much weaker than expected \cite{Coles2006}.

In this paper, we report a type of short radio burst observed with the LOw Frequency ARray (LOFAR \cite{vanHaarlem2013}) --- a spike-like repeating burst pair that can appear either isolated or in chains where each fine structure is repeated at the same frequency with a delay of about 4\,s. These repeating burst pairs are characterized by low drift rates and narrow spectral widths of tens of kilohertz. The repeated component is often fainter, with a reduced drift rate and spatially separated. Using the statistical results on the temporal, spectral, and spatial characteristics, we suggest a formation mechanism in the context of anisotropic scattering/reflection of the harmonic radio emission. The observation of the naturally occurring echo (scattered harmonic emission) provides insights as to how and where the radio signal is reflected, suggesting anisotropic scattering weakens the radar signal, potentially opening a new avenue for radio wave scattering experiments.

\section*{Results}
\subsection*{The Discovery of Repeating Burst Pairs} 
The radio data presented in this paper are LOFAR observations from the Low Band Antenna in the outer configuration with a baseline of about 3.6~km, covering the frequency range of 30 to 80 MHz, which probes the mid corona. 
The LOFAR beam-formed data have a temporal and spectral resolution of 10 ms and 12.2 kHz, respectively, with the beam-formed images constructed from 217 interferometrically synthesized beams using the same method as described in \cite{Kontar2017}. The beam pointings are centered on the Sun and extend to 3~R$_\odot$. The flux was calibrated to solar flux units (sfu; 1 sfu = $10^{-22}$~W m$^{-2}$ Hz$^{-1}$) with uncertainty about 1~sfu using observations of Tau A \cite{Kontar2017}.
During 9 July 2017, a type of radio burst was detected co-spatially along with numerous mini jets originating from the active region located at the sky-plane coordinates ($-383\arcsec,-156\arcsec$) as seen in extreme ultraviolet (EUV) passbands of the Atmospheric Imaging Assembly (AIA; \cite{Lemen2012}) onboard the Solar Dynamics Observatory (SDO; \cite{Pesnell2012}).
Data from the Helioseismic and Magnetic Imager (HMI; \cite{Schou2012}) onboard SDO, as well as a potential-field source-surface extrapolation (PFSS; \cite{Schatten1969, Altschuler1969}) are used to describe the surrounding magnetic environment.

\begin{figure}[!ht]
\centering
\includegraphics[width=1\linewidth]{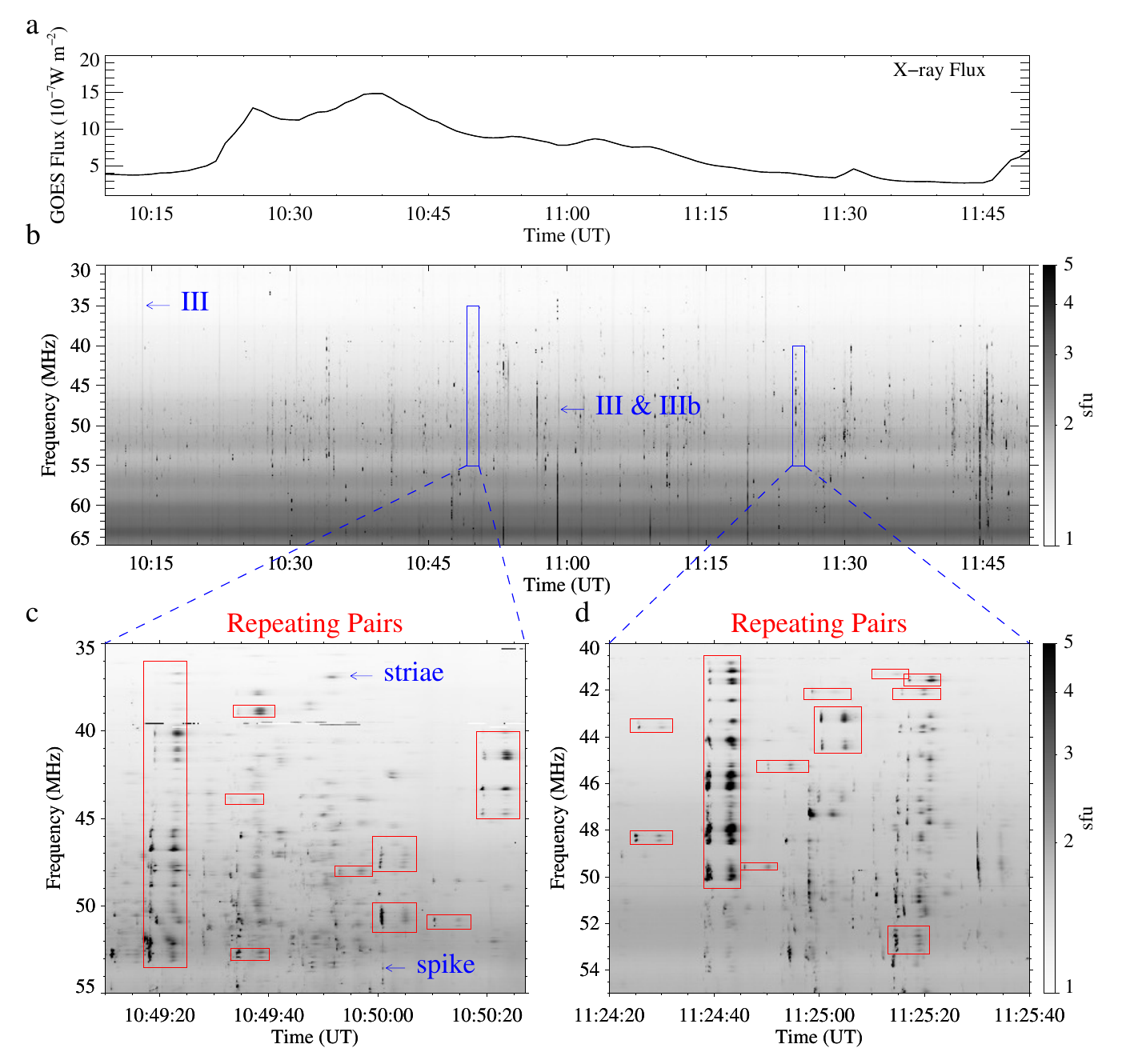}
\caption{\textbf{Radio dynamic spectra showing various burst types during the event.} \textbf{a} The X-ray flux from the GOES spacecraft.
\textbf{b} The dynamic spectrum between 10:10 and 11:50 UT on 9 July 2017. The blue rectangles in the upper panel mark the zoomed-in spectra shown in the lower panels. \textbf{c and d} Zoomed-in dynamic spectra from 10:49:10 to 10:50:27 UT and from 11:24:20 to 11:25:40 UT displaying repeating burst chains and repeating isolated bursts. The red boxes indicate examples of repeating burst pairs(RPs).}
\label{fig:fig1}
\end{figure}

The black curve in Figure~\ref{fig:fig1}a shows the GOES soft X-ray flux.
The dynamic spectrum from LOFAR, covering 30–65 MHz, is shown in Figure~\ref{fig:fig1}b.
In the dynamic spectra, the y-axis is arranged from high to low frequency, which corresponds to increasing altitude in the solar corona. This orientation reflects the inverse relationship between plasma frequency and coronal height. Throughout the observing period, numerous broadband emissions such as Type III and IIIb bursts were observed, as well as many short timescale and narrowband fine structures.

A central feature observed during this event is the repeating spike-like fine structures (Figure \ref{fig:fig1}c,d), which manifest as either isolated bursts or as chains occurring simultaneously across different frequencies.
Many of the repeating bursts form prominent chains, with each repeating spike occurring at the same frequency as the preceding one, resulting in two parallel columns separated in time by approximately 4 seconds. The initial bursts in these chains have shorter durations and higher intensities, while the subsequent bursts in the chains exhibit longer durations and lower intensities. We also discovered numerous repeating pairs that appeared in shorter chains (with a smaller frequency extent) or in isolation. 
Based on these observations, we define \textbf{a repeating burst pair} as two independent pulses with different lifetimes and intensities, occurring at the same frequency and separated by approximately four seconds. In each pair, we designate the first pulse as the earlier component (E) and the subsequent pulse as the delayed component (D).

\subsection*{Repeating Burst Pairs Characteristics}
\begin{figure}[htbp]
\centering
\includegraphics[width=0.95\linewidth]{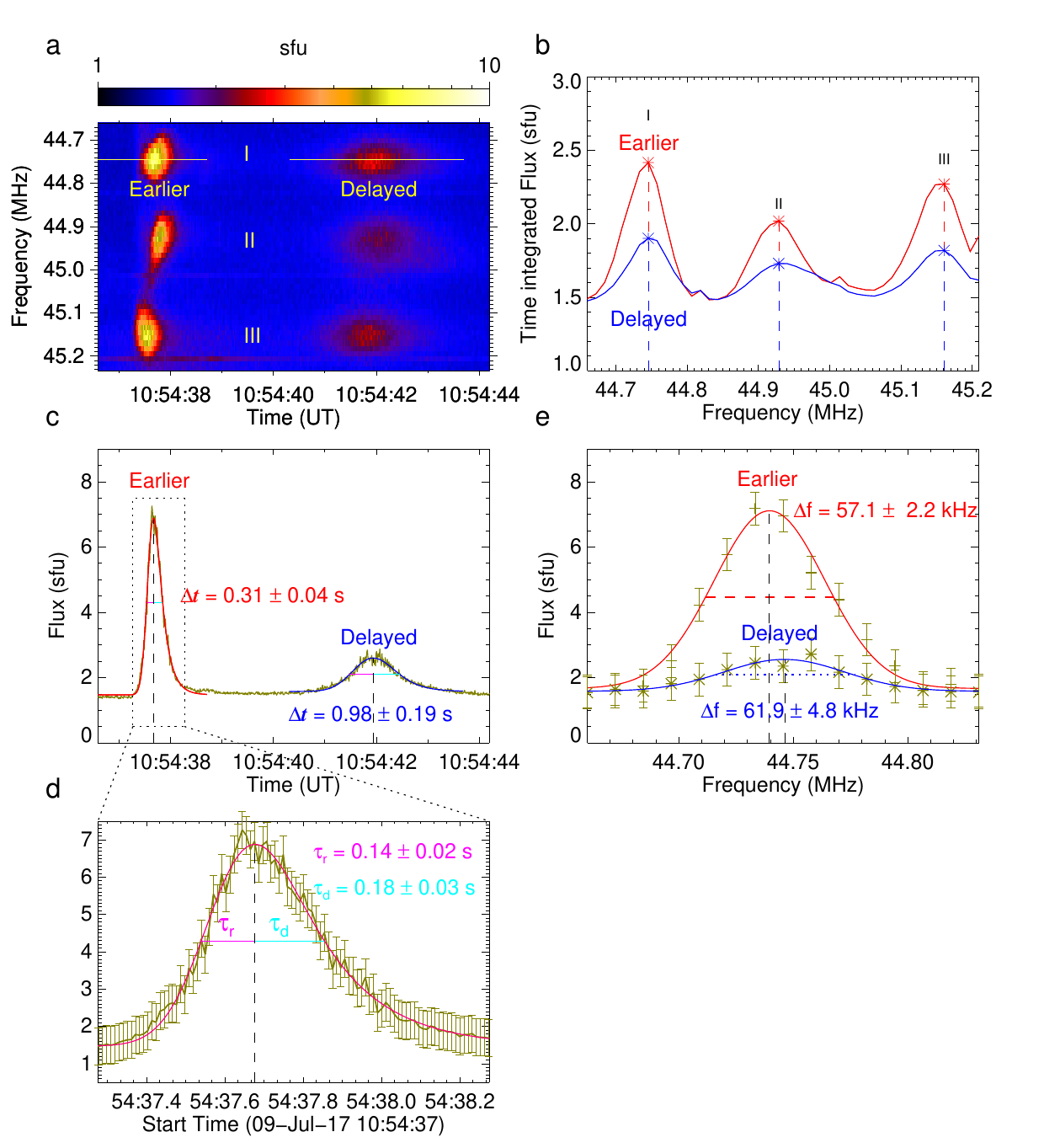}
\caption{\textbf{Dynamic spectrum and flux profiles of three repeating burst pairs (RPs)}. \textbf{a} Radio dynamic spectrum showing three RPs: I, II and III. The horizontal lines represent both the time-integration interval for the flux in panel \textbf{b} and \textbf{c}. Time-integrated flux against frequency over the time intervals corresponding to the earlier (red) and delayed (blue) bursts. \textbf{c} Time profile along the central frequency of the bursts at 44.74~MHz. The olive curve shows the observed flux, and the red and blue curves display the asymmetrical Gaussian fits to each component. The dashed vertical lines show the fitted Gaussian peak time. The magenta and turquoise horizontal lines denote the fitted Half-Width at Half-Maximum (HWHM) durations. \textbf{d} The frequency–flux profiles and Gaussian fits for the earlier (live plus symbols and red curve) and delayed (olive star symbols and blue curve) RP I at peak times. The vertical lines indicate the frequencies at which the fitted Gaussian peaks for each burst. The horizontal dotted lines mark the fitted Full-width at Half-Maximum (FWHM) spectral widths. \textbf{e} A zoomed-in profile of the earlier burst indicated by the dashed box in \textbf{c}. The error bars in panels b and e represent the measured error of flux.}
\label{fig:fig2}
\end{figure}

Figure~\ref{fig:fig2} shows an example of the characteristic measurements from repeating pairs in a chain. 

Figure \ref{fig:fig2}a presents the dynamic spectrum composed of three of the RPs: I, II, and III. The horizontal lines mark both the time-integration interval for the integrated flux shown in Figure 2b and the analysis frequency of the time-flux profile presented in Figure 2c. Figure~\ref{fig:fig2}b shows the time-integrated flux of each of the components in the chain, computed over time ranges encompassing the E and D chains as indicated by the horizontal lines in Figure \ref{fig:fig2}a. The central frequency of each burst is defined as the frequency at which the integrated flux reaches its maximum. A comparison of the flux profiles from the E and D components shows clear alignment between the time-integrated central frequencies. Additionally, the E components are brighter than the delayed ones.

The time-flux profile of repeating pair I at its central frequency of 44.74 MHz is shown in Figure~\ref{fig:fig2}c and the frequency flux profiles of each
burst component at the time of peak flux $I_0$ are presented in Figure~\ref{fig:fig2}d. The olive curve represents the measured flux, while the red and blue curves show the asymmetrical Gaussian fits to the flux profiles of the E and D components, respectively. Given the Gaussian rise and exponential decay of the time profiles, we have adopted an asymmetrical Gaussian fitting approach (see Methods subsection Repeating Pair Screening and Spectral Parameter Measurements).

The dashed vertical lines in Figure~\ref{fig:fig2}c indicate the time when the flux of E and D bursts achieve their respective peaks at $(6.7\pm1)$~sfu and $(2.5\pm1)$~sfu, respectively, such that the E component is a factor of $2.6\times$ brighter. The time separation between two peaks, denoted as $\Delta{t_s}$, is about 4~s. Figure~\ref{fig:fig2}e displays the zoomed time-flux plot of the E component indicated by the dashed box in  Figure~\ref{fig:fig2}c. The magenta and turquoise horizontal lines represent the half-width at half-maximum (HWHM) of the duration during the rise ($\tau_r$) and decay ($\tau_d$) phase, respectively. The full-width at half-maximum (FWHM) duration $\Delta{t}$ of each component is then $\Delta{t}=\tau_r+\tau_d$, yielding values of about 0.32~s and 1.0~s for the E and D components, respectively. The frequency flux profiles of each burst component at the time of peak flux $I_0$ are presented in panel d and are fitted with symmetric Gaussian profiles. The horizontal dotted lines represent the spectral FWHM bandwidth of about 60~kHz for both components.

\begin{figure}[htbp]
\centering
\includegraphics[width=1\linewidth]{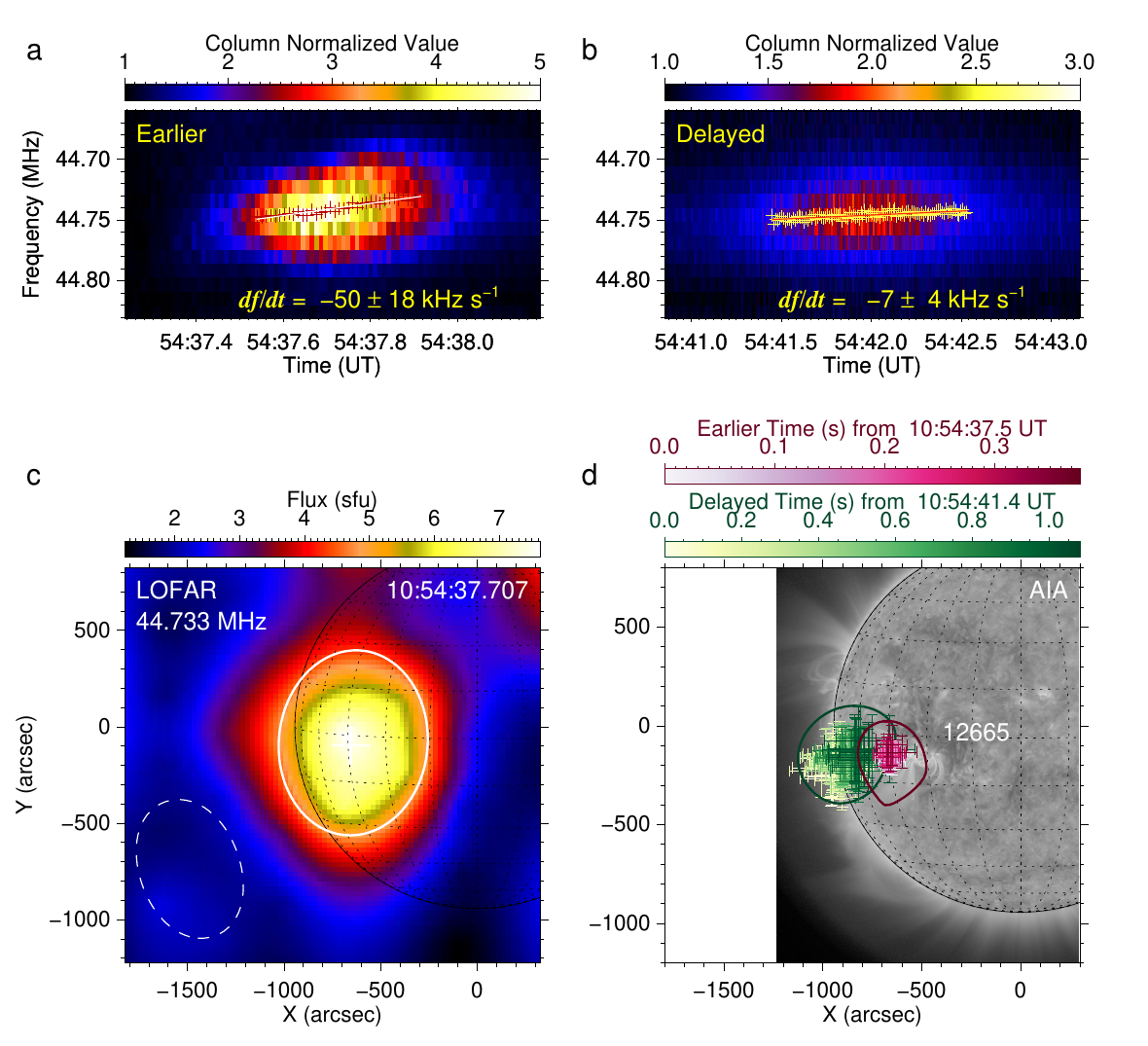}
\caption{\textbf{Zoomed-in dynamic spectrum and centroids of repeating pair (RP) I.} \textbf{a} and \textbf{b} Dynamic spectrum of the E and D component overlaid with the measurement of the frequency drift rates. The red/yellow pluses show the peaks of each Gaussian fit to the flux profiles at each time bin, and the yellow/red lines show the linear fit to derive the frequency drift rate. \textbf{c} LOFAR image at the peak intensity of the earlier burst at 44.73 MHz at 10:54:37.7 UT. The white contour marks the 2D Gaussian fit at the FWHM level, with the centroid marked by the white cross. The dashed white oval indicates the half-maximum synthesized LOFAR beam. \textbf{d} Fitted source centroids the radio images of the E and D components at the time and frequencies indicated with red pluses in Figure \ref{fig:fig3} a and b overlaid on the AIA image at 171 $\text{\AA}$. 
The error bars along the $x$ and $y$ axes represent the standard error estimated via error propagation. 
The color gradient from light to dark indicates a temporal progression. The red and green contours indicate 90\% intensity levels of E and D bursts at peak time.}
\label{fig:fig3}
\end{figure}

Figure \ref{fig:fig3}(a),(b) displays the zoomed-in dynamic spectra of burst I overlaid with the resulting fit to determine the frequency drift rate $\mathrm{d}f/\mathrm{d}t$ of each component. 

To calculate the frequency drift rate of each burst, we first determine the average flux over the frequencies between two frequency-flux minima and fit it with a Gaussian function at each time across the FWHM duration of the burst. The frequency drift rate is derived by performing a linear fit to the time and frequency peaks. The E component has a frequency drift rate of $\mathrm{d}f/\mathrm{d}t=(-50\pm18)$~kHz s$^{-1}$, and the delayed burst has a comparatively reduced drift rate of $\mathrm{d}f/\mathrm{d}t=(-7\pm4)$~kHz s$^{-1}$ (the errors here represent one-sigma linear fitting errors); a factor of 7 lower. The drift rate of the E component is comparable to that of spikes at a few tens of kilohertz per second \cite{Clarkson2023} but lower than that of drift pairs (2-8~MHz s$^{-1}$) \cite{Kuznetsov2019} and Type III bursts (~1-20~MHz s$^{-1}$) \cite{Reid2018} at the same frequency. The negative value likely indicates that the burst source is moving away from the Sun.

Following the approach in \cite{Kontar2017}, we use 2D Gaussian fitting on LOFAR images to calculate the source centroids and the source area. As mentioned in  \cite{Condon1997, Kontar2017}, the ellipse centroid position is determined to an accuracy significantly better than the angular resolution of a single beam measurement.
Due to the low brightness of some bursts and the presence of sidelobes, the 2D Gaussian fitting may not always converge.
In this investigation, the field-of-view (FOV) is restricted to a range of $-1800\arcsec$ to $300\arcsec$ along the x-axis and $-1200\arcsec$ to $800\arcsec$ along the y-axis, as shown in Figure~\ref{fig:fig3}c, and encompasses the primary lobe of each burst. The fitted centroid positions are then corrected for the average effects of ionospheric refraction as in \cite{Gordovskyy2022, Clarkson2023}. The 2D Gaussian fit at the peak time of the earlier burst is represented by the white contour at the FWHM level in Figure~\ref{fig:fig3}c, with the white plus indicating the centroid. The fit covers an area of approximately $(160\pm20)$~arcmin$^{2}$ (the error denotes one-sigma fitting uncertainty), which coincides with the FWHM area of spikes reported in \cite{Clarkson2021}.

The image presented in Figure \ref{fig:fig3}d showcases the fitted centroids with error bars of the radio images for both the E and D bursts at different times overlaid on the AIA 171 $\text{\AA}$ image. 
The color depth illustrates the extent of the time increase.
The centroids of the earlier burst are concentrated closer to the active region and centralized, whilst those of the delayed burst appear to be distributed farther away from the center of the active region and spread over a larger distance in the projected sky-plane. The colored contours map the 90$\%$ intensity level at the peak time of the bursts. At the peak times of each burst component, the centroids are displaced by about 270~arcsec. The largest distance between the centroids of the E and D components is around 430~arcsec. 

\subsection*{Statistics of Repeating Burst Pairs}
\begin{figure}[!ht]
\centering
\includegraphics[angle=0, width=1.\linewidth]{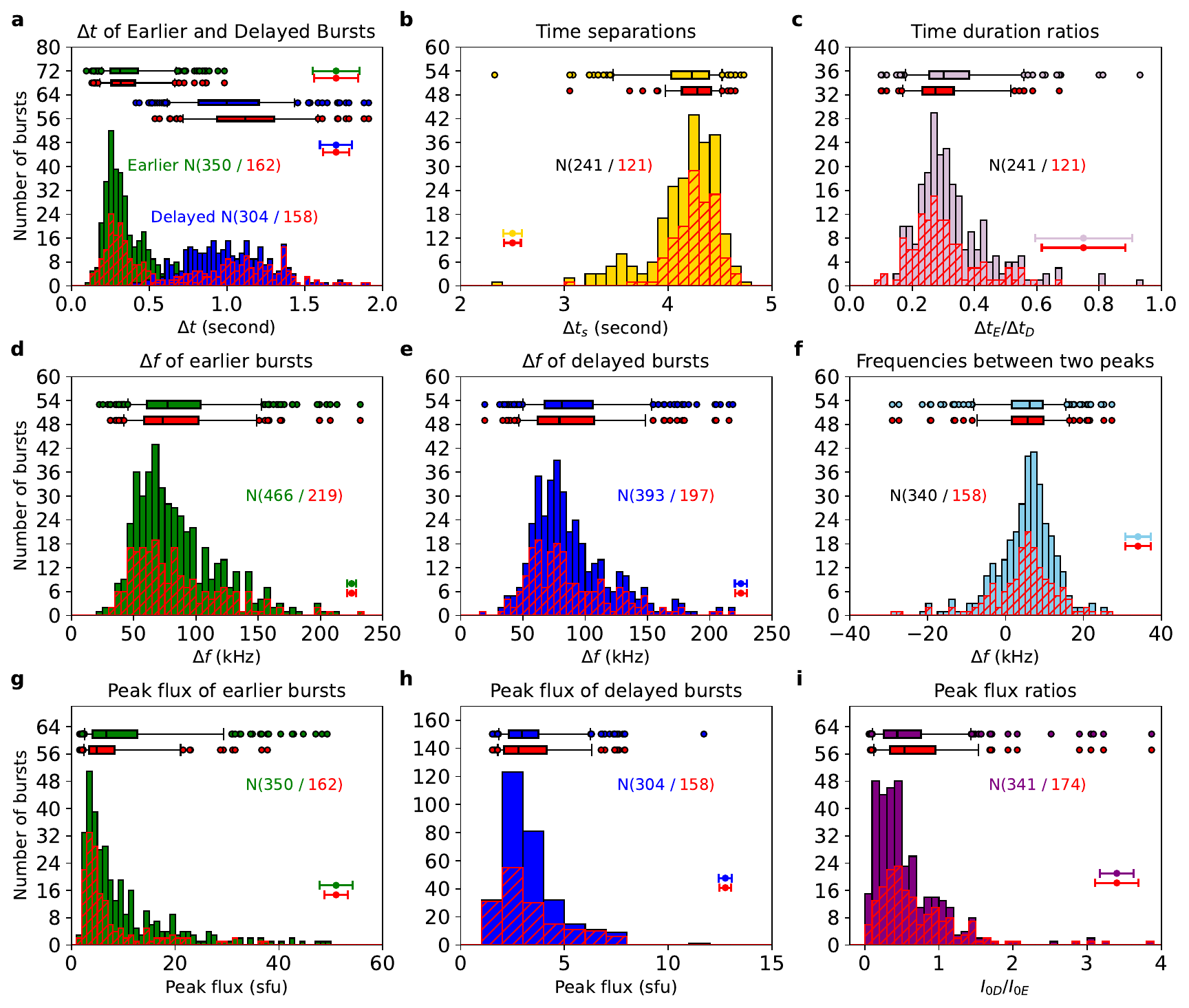}
\caption{\textbf{Statistical results of the burst pair characteristics covering a frequency range of 30 to 60 MHz}. \textbf{a} Burst duration at FWHM for the earlier (green) and delayed (blue) components.
\textbf{b} Time interval between E and D bursts of each pair.
\textbf{c} Duration ratio of each pair ($\Delta t_E / \Delta t_D$).
\textbf{d–e} Frequency bandwidth at FWHM for E and D bursts.
\textbf{f} Difference in central frequency between the two components.
\textbf{g–h} Peak flux of E and D components.
\textbf{i} Peak flux ratio ($(I_0)_D / (I_0)_E$) of each pair.
Sections with red diagonal lines superimposed on the bars denote the data used for imaging.
Values in parentheses give the number of bursts per characteristic for which the Gaussian fit converged:
the first number corresponds to bursts measured across the full frequency range, and the second to those below 48 MHz that were included in the imaging analysis.
Box plots show the median (middle line), the 25th and 75th percentiles (box), the 5th and 95th percentiles (whiskers), and outliers (single points) for the relevant statistical parameters of total bursts (green, blue, gold, thistle, skyblue, or purple) and imaging bursts (red). The error bars represent the median of the standard deviations obtained from the Gaussian fits for each parameter.}
\label{fig:4}
\end{figure}

From over one thousand individual pairs identified from the two-hour observation period shown in Figure~\ref{fig:fig1}, 613 pairs were chosen for analysis in dynamic spectra, where bursts with overlapping structures were omitted.We restrict imaging to bursts occurring at frequencies less than approximately 48~MHz, a range consistent with that adopted by \cite{Gordovskyy2019}, given the negligible frequency‑dependence of the main lobe's position. As such, imaging analysis is conducted using 309 burst pairs. The statistical results obtained from dynamic spectra are presented in Figure~\ref{fig:4} with the bursts that were included in the imaging analysis highlighted in red. We note that data is not included for a specific characteristic where the fitting procedure failed to converge, and the remaining number of measurements for each characteristic is shown in parentheses.

The box-and-whisker plots in Figure~\ref{fig:4}(a–k) illustrate the distribution of the relevant statistical parameters for total bursts and for bursts with imaging, depicting the median (middle line), the 25th and 75th percentiles (box), the 5th and 95th percentiles (whiskers), and outliers (single points). The relevant statistical parameters are also listed in Supplementary Table 1. Colors denote sample groups: green (E component), blue (D component), and black, light blue, and purple (differences or ratios between components). Red indicates samples for which imaging data are available.

\textbf{Time.} Figure \ref{fig:4}a indicates the distributions of the FWHM duration for E and D bursts. 
The E component durations range from $0.14$~s to $1.10$~s, with the majority clustered near $0.25$~s to $0.43$~s and a median of $0.31$~s, while the durations of the D components range from $0.33$~s to about 1.82~s with a majority between $0.82$ to $1.2$~s and a median of $1.0$. Figure \ref{fig:4}b displays the distribution of the time separation $\Delta{t_s}$ between two intensity peaks of each burst pair. The distribution shows two peaks: a weaker peak near $3.5$~s and a stronger peak near $4.2$~s. Figure \ref{fig:4}c shows the ratio of earlier to delayed duration $\Delta{t_E}/\Delta{t_D}$ with a spread from $0.22$ to $0.38$ and a peak at around $0.3$. 

\textbf{Frequency.} The bandwidth distribution of the earlier bursts in Figure~\ref{fig:4}d is similar to that of the delayed bursts in Figure~\ref{fig:4}e, with a slight difference. The spectral FWHM of earlier bursts median at about 70\,kHz, while that of the delayed one peaked at about 80\,kHz. The distribution of the central frequency difference between each component is displayed in Figure~\ref{fig:4}f with a median near 8\,kHz. Given the 12.2\,kHz LOFAR spectral resolution, this suggests that in most cases the E and D bursts in a pair occurred at consistent frequencies.

\textbf{Peak flux.}
Figure~\ref{fig:4}(g--h) presents the peak flux \( \mathrm{I}_{0} \) distributions of the E and D bursts. 
The E components span a range of peak fluxes up to \(\sim45\)\,\text{sfu}, with the majority concentrated between 3 and 6\,\text{sfu}. 
In comparison, the D components are generally weaker and exhibit a narrower distribution, with peak fluxes reaching up to \(\sim8\)\,\text{sfu} and a median near 3\,\text{sfu}.
The peak flux ratios \( \mathrm{I}_{0\mathrm{D}}/\mathrm{I}_{0\mathrm{E}} \) shown in Figure~\ref{fig:4}(i) range from 0.1 to 3.4, with lower and upper quartiles of 0.3 and 0.8, respectively. The distribution exhibits a peak near 0.3 and a median of approximately 0.5. 
The mean uncertainty is approximately 1\,\text{sfu} for the peak flux and 0.20 for the peak flux ratio. These values indicate that, within a given RP, the E component is typically two to three times brighter than its delayed counterpart.

\begin{figure}[!ht]
\centering
\includegraphics[angle=0, width=1.\linewidth]{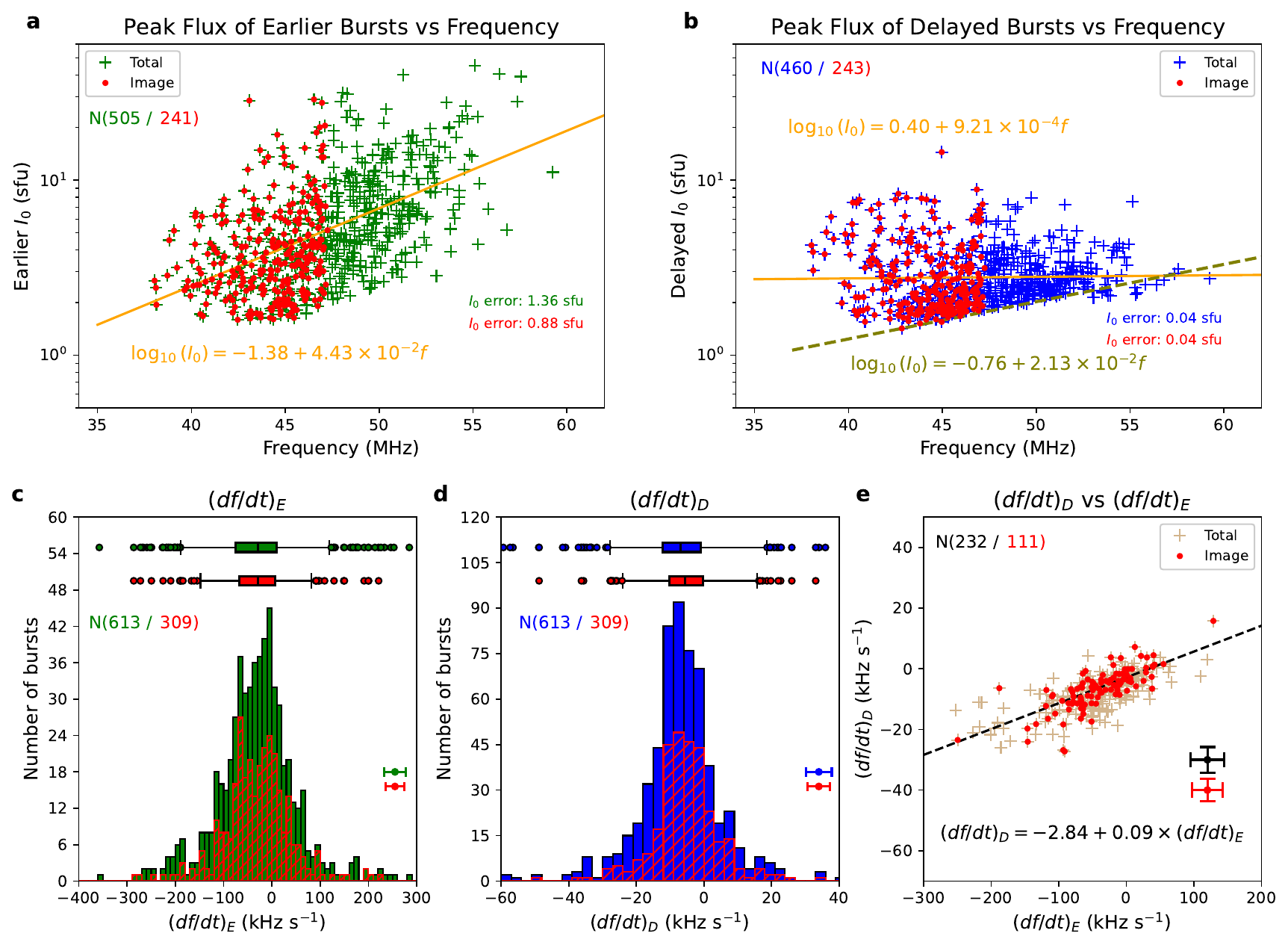}
\caption{\textbf{Statistical parameters of repeating pairs (RPs) and the relationship between them.} Panels \textbf{a} and \textbf{b} show the relationship between peak flux and frequency for the earlier (E) and delayed (D) components, respectively. Small red circles denote the samples that are accompanied by images. Orange lines and their corresponding equations represent linear fits to the data. The olive line in panel \textbf{b} indicates the visually estimated trend of the minimum peak flux as a function of frequency. \textbf{c-d} Frequency drift rate $\mathrm{d}f/\mathrm{d}t$ of E and D components, respectively. \textbf{e} Scatter plot of $(\mathrm{d}f/\mathrm{d}t)_D$ vs. $(\mathrm{d}f/\mathrm{d}t)_E$ for the RPs. The values in parentheses show the number of measurements for each characteristic where the fit converged. The first number corresponds to measurements made across the full frequency range, while the second number (if present) corresponds to those below 48 MHz that were included in the imaging analysis. The box and whisker plots show the sample minimum, lower quartile, median, upper quartile, and sample maximum of the relevant statistical parameter. Colors correspond to data groups: earlier (green), delayed (blue). The subset that includes imaging data is shown in red. The error bars or ${I_0}$ error indicated the median error of each parameter. }
\label{fig:5}
\end{figure}

Given that the distribution of the peak flux density of the radio sources deviates substantially from a normal distribution, we further examined the relationship between peak flux density and source central frequency. 
The results are presented in Figures~\ref{fig:5}(a) and ~\ref{fig:5}(b) for the E and D components, respectively. 

Among the earlier bursts, the peak flux exhibits a weak increasing trend with frequency, $\log_{10}(I_{0E}) = -1.38 + 4.4 \times 10^{-2} f$; however, no such trend is evident in the D components. The slope of the linear fitting for the D components, $\log_{10}(I_{0D}) = 0.402 + 9.2 \times 10^{-4} f$, is much smaller, approximately $9.2 \times 10^{-4}$. Nevertheless, the visually estimated trend of the minimum peak flux shows an increasing trend with frequency, which can be described as $\log_{10}(I_{0D}) = -0.76 + 2.13 \times 10^{-2} f$.

\textbf{Frequency drift rates.} The frequency drift rates \(\mathrm{d}f/\mathrm{d}t\) (Figure~\ref{fig:5}c) of the E components showed significant variation, ranging mainly from \(-357\)\,kHz\,s\(^{-1}\) to \(246\)\,kHz\,s\(^{-1}\) with a peak value around \(-50\)\,kHz\,s\(^{-1}\) and a median value of \(-30\)\,kHz\,s\(^{-1}\). The delayed bursts exhibited a significantly narrower distribution from approximately \(-48\)\,kHz\,s\(^{-1}\) to \(18\)\,kHz\,s\(^{-1}\), with the majority clustered around \(-13\)\,kHz\,s\(^{-1}\) to \(-3\)\,kHz\,s\(^{-1}\) and a median of \(-7\)\,kHz\,s\(^{-1}\) (Figure~\ref{fig:5}d). The mean frequency drift rate error (one-sigma linear fitting uncertainties) is around 24\,kHz\,s\(^{-1}\) for the E component and 4\,kHz\,s\(^{-1}\) for the D component. The majority of all bursts present a negative frequency drift. The earlier burst typically exhibits a stronger frequency drift rate than its corresponding D component, although the two usually share the same sign of drift, indicating similar but weaker spectral evolution in the delayed burst.

Figure~\ref{fig:5}e displays the relationship between the frequency drift rate of the D component and that of the E component. The data are described by a linear fit with a slope of approximately 0.07.

\begin{figure}[htbp]
\centering
\includegraphics[width=1\linewidth]{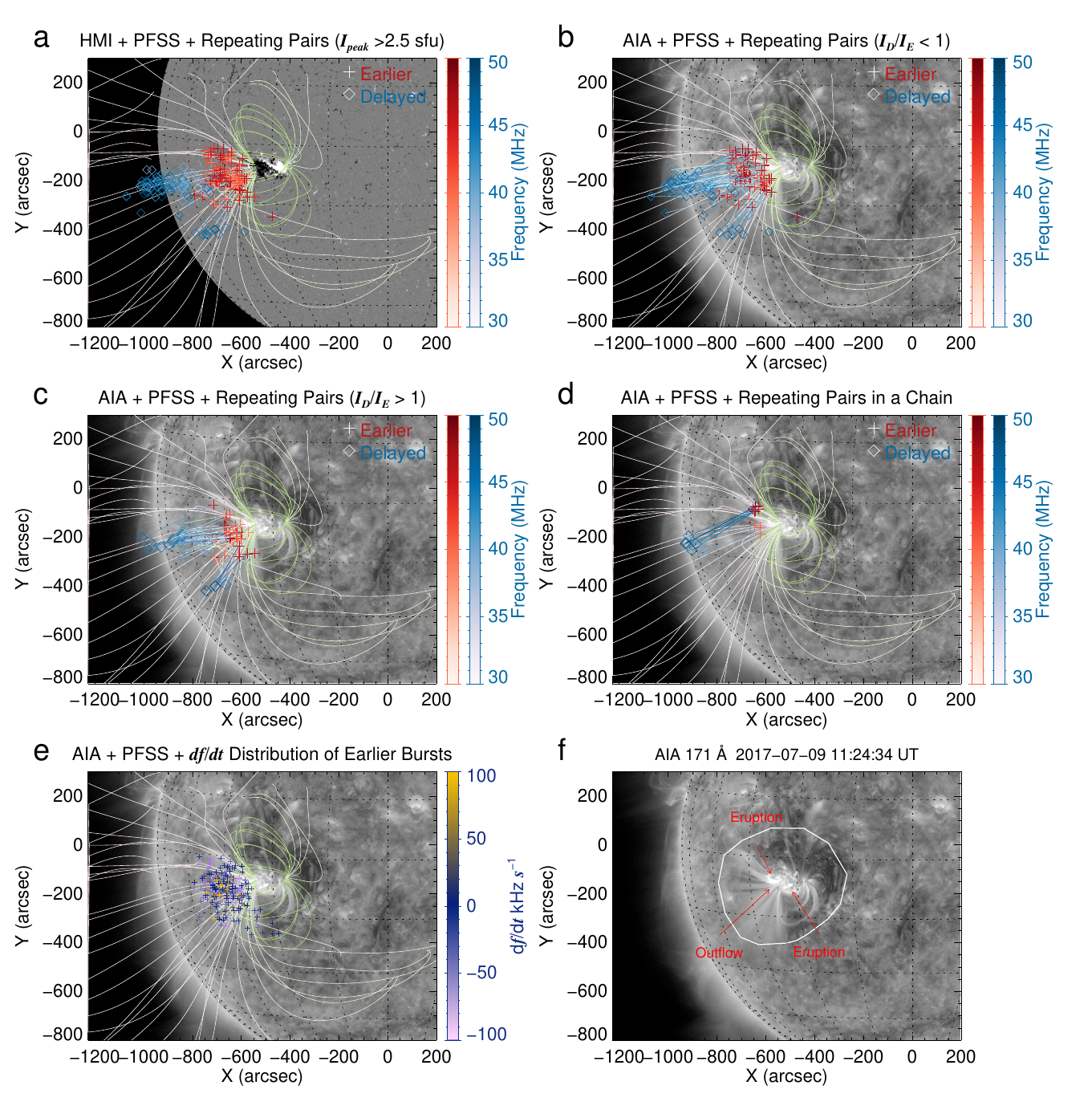}
\caption{\textbf{Centroids of each burst component overlaid on the surrounding magnetic environment.} The background of panel \textbf{a} shows the HMI magnetogram, while panels \textbf{b-f} show the AIA 171 \AA~image. The light pink and green curves represent magnetic field lines from a PFSS extrapolation, corresponding to open and closed field lines, respectively. The red pluses and blue diamonds in panels \textbf{a-d} mark the centroid positions of the E and D components, respectively, with the color gradients corresponding to different frequencies as indicated by the colorbar within each panel. \textbf{a} Centroids of the earlier bursts with peak flux $>2.5$~sfu (red pluses) and the D components with peak flux $>2.5$~sfu (blue diamonds). \textbf{b} Centroids of burst pairs with flux ratio $<1$ and peak fluxes of both components $>2.5$~sfu. Each pair is connected by a light yellow or light blue line. \textbf{c} Similar to panel \textbf{b} but for burst pairs with flux ratio $\geq 1$. \textbf{d} Several repeating burst pairs (RPs) from the same chain, as highlighted by the red box in Figure~\ref{fig:fig1}d. \textbf{e} Centroids of earlier bursts, color-coded according to drift rate. \textbf{f} AIA 171 \AA~image at 11:24:34 UT showing the morphology and eruptions. Red arrows indicate eruptions and outflow. }
\label{fig:fig_centroid}
\end{figure}

\textbf{Centroid.} Figure \ref{fig:fig_centroid} presents the spatial distribution of the E and D components of repeating pairs relative to the surrounding magnetic environment.

The coronal magnetic field was extrapolated using the PFSS package \citep{Schrijver+2003} in SolarSoftWare (SSW) over the active region of interest, starting at 1.2~R$_\odot$ above the photosphere with a source height at 2.5~R$_\odot$.
The centroid positions of the E and D components are shown in Figure \ref{fig:fig_centroid}a, where only components with peak flux $>2.5$ sfu are included to mitigate source overlap. The same flux threshold is used in Figures \ref{fig:fig_centroid}b and \ref{fig:fig_centroid}c.
The centroid distribution of each burst pair, connected by a line, is illustrated in Figures \ref{fig:fig_centroid}b and \ref{fig:fig_centroid}c, with pairs having flux ratios $<1$ and $\geq1$ displayed separately in each panel.
Figure~\ref{fig:fig_centroid}e illustrates the frequency drift rate distribution of the earlier bursts, while Figure~\ref{fig:fig_centroid}f shows the active region as imaged by AIA at 171 \AA.

The centroid distributions presented in Figure \ref{fig:fig_centroid}a indicate that earlier bursts (red plus signs) are preferentially located near the negative polarity magnetic field of the active region, whereas D components (blue diamonds) exhibit a more scattered distribution and are situated at greater distances. The light-to-dark color gradient corresponds to increasing frequency from 30 to 50 MHz. It should be noted that the centroid positions represent projected locations on the plane of the sky; consequently, the heliocentric distance cannot be inferred from the centroids alone. Under the assumption that plasma density decreases monotonically with height above the solar surface, lower-frequency emission is expected to originate at higher altitudes than higher-frequency emission. Based on this premise, the observed frequency distribution implies that the bursts occupy an extended spatial volume. Considering the topology of the extrapolated magnetic field, all bursts appear to originate from a cone-shaped magnetic structure arching over the negative polarity region, characterized by closed field lines on the western side and open field lines on the eastern side.
Based on Figures \ref{fig:fig_centroid}b and \ref{fig:fig_centroid}c, the displacement between the E and D source centroids in each pair ranges from a few tens of arcseconds to about 400$\arcsec$ in the sky plane. No significant difference in the spatial distribution is observed between pairs with different flux ratios.
The lines connecting the E and D components within each pair can be broadly divided into two groups: one oriented approximately west–east and the other northwest–southeast. It is worth noting that magnetic field lines represent the direction, not the strength, of the magnetic field (in this paper). In both groups, the direction of the connecting line is generally parallel to that of the magnetic field lines. 
Figure~\ref{fig:fig_centroid}d shows the centroid positions of repeating bursts forming an elongated chain during the interval 11:24:37–11:24:45 UT, over a frequency range of 40.5–50.5 MHz (indicated by the red box in Figure~\ref{fig:fig1}d). This chain exhibits a bulk frequency drift similar to that of type III bursts. The centroids of the E components, denoted by plus symbols, are located near the right boundary of the funnel-shaped structure and are aligned approximately north–south (likely corresponding to increasing altitude) as the central frequency decreases (from dark to light red). 
The centroids of the D components are situated near the left boundary of the funnel, with higher-frequency emissions located slightly farther outward than lower-frequency ones.

Figure~\ref{fig:fig_centroid}e shows the distribution of the frequency drift rate of the earlier bursts, with pink indicating large negative drift rates, yellow representing large positive drift rates, and blue corresponding to small drift rates near zero. The results suggest that, in the central region of the funnel, the distribution appears to be approximately random. However, it should be noted that this finding may be significantly influenced by projection effects. Interestingly, on the far right side of the cone—where the magnetic field lines are predominantly closed—the centroids are almost exclusively blue, indicating consistently small drift rates.

\begin{figure}[!ht]
\centering
\includegraphics[width=1\linewidth]{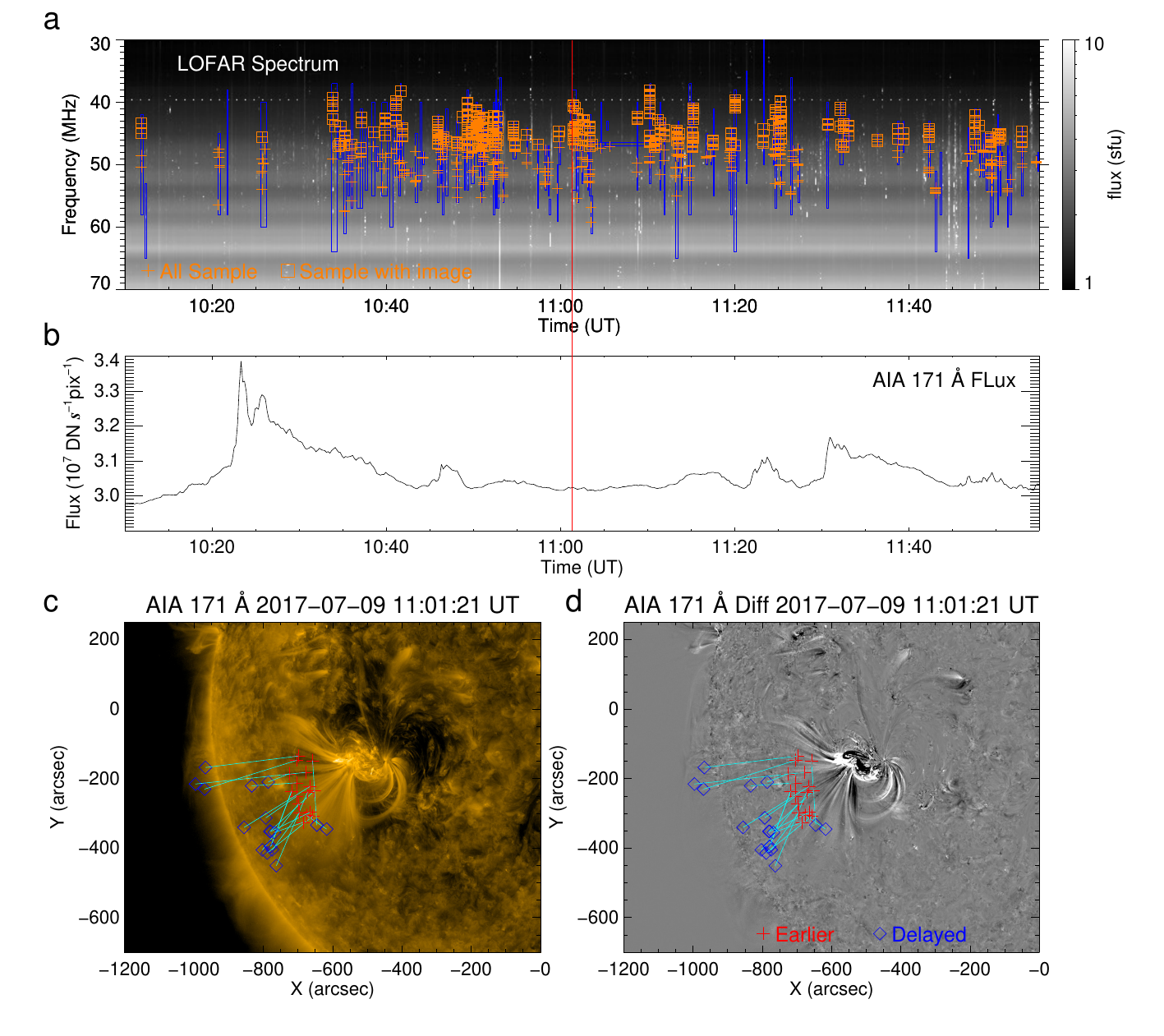}
\caption{\textbf{Temporal evolution of solar eruptive activity and associated radio bursts}. \textbf{a} Dynamic spectrum observed by LOFAR. Blue slices/boxes indicate the initial sampling range; orange plus symbols denote secondary selection of distinct repeating burst pairs (RPs) across all frequencies; orange squares mark sources selected for imaging. \textbf{b} AIA time-density flux profile integrated over the active region. The red vertical line marks the time of the AIA images presented in the panels \textbf{c} and \textbf{d}, where the left and right panels show the original and base-difference images, respectively. The red plus and blue diamond (connected by a turquoise line) indicate the centroids of the earlier and delayed components of a pair. The accompanying animation is available as Supplementary Movie 1.}
\label{fig:fig_timebar}
\end{figure}

Figure~\ref{fig:fig_centroid}f presents an AIA 171\AA\ image taken at 11:24:34~UT, showing the magnetic morphology associated with the RPs. Numerous eruptions are continuously observed to initiate from the core of the active region, as indicated by the red arrows labeled ``eruption''. Simultaneously, multiple faint outflows or jets continually emanate from the root of the cone-shaped structure, marked by the red arrow labeled ``outflow''. 
These eruptions and outflows may be the trigger for the generation of the RPs. However, establishing a straightforward relationship between these phenomena is challenging, given the 12~s cadence and $0.6\arcsec$ spatial resolution of AIA, and the sub-second duration (generally less than 1~s) of the repeating bursts themselves. The white closed curve encloses the region over which the flux is calculated, and the results are presented in Figures~\ref{fig:fig1} and \ref{fig:fig_timebar} for comparison with other results. 

Nevertheless, investigating the statistical correlation between repeated eruptions and the overall eruptive activity of active regions could be a step toward understanding the underlying mechanisms. Figure~\ref{fig:fig_timebar} shows the temporal evolution of radio emission based on LOFAR (top) and the EUV density flux of the active region based on AIA 171 \AA\ (middle). 

The accompanying animation (~Supplementary Movie 1.mp4~) shows the spatiotemporal emergence of repeating pairs as the active region evolves (12 s cadence). A red plus and blue diamond connected by a turquoise line mark the centroids of the E and D components of each pair. No strong radio emission is observed when the EUV flux reaches its peaks, and conversely, when the radio emission intensifies, the EUV flux is relatively weak. This lack of temporal correspondence can be attributed to the different source regions: the EUV flux originates primarily from the lower-lying flare region, while the radio sources of the repeating pairs are located higher in the corona. Furthermore, the lifetime of a single burst is less than 1 second, whereas the AIA observations have a cadence of 12 seconds. Future observations of the higher corona with improved temporal and spatial resolution may help address this limitation.

From the statistical analysis of the burst pairs, we find that:
\begin{itemize}
    \item The peak fluxes of most bursts are lower than 5 sfu, with the brightest bursts near 45~sfu. The D components tend to have a weaker peak flux than the E component, with the majority of the pairs concentrated around a flux ratio of \( \mathrm{I}_{0D}/\mathrm{I}_{0E}\sim0.5 \). 
    \item The FWHM duration of E components ranges from 0.14 to $1.1$~s with the majority clustered near 0.30~s, while the duration of the D components ranges from 0.3 to 1.82~s with a peak between (0.8--1.2)~s. The average time separation (delay) between the pairs is about 4.3~s.
    \item The bandwidth distribution of E and D components are similar, with a median at (70--80)~kHz. The components of each burst pair occurred at the same frequency within the spectral resolution of 12~kHz.   
    \item The frequency drift rates of the E components varied widely, ranging between approximately $-357$~kHz s$^{-1}$ to $246$~kHz s$^{-1}$ with a peak at $-50$~kHz s$^{-1}$. In contrast, the delayed bursts displayed a narrower distribution, ranging approximately from $-48$~kHz s$^{-1}$ to $20$~kHz s$^{-1}$ with a peak at $-10$~kHz s$^{-1}$. 
    In the majority of RPs, the frequency drift rate of the D component shares the same sign as that of the E component, indicating similar source movements in most cases.    
    \item The earlier burst centroids are primarily concentrated in an area closer to the negative polarity magnetic field of the active region, whilst the D component centroids appear farther away from the active region and spread across a larger area. The displacement between the E and D source centroids in each pair ranges from a few tens of arcseconds to 400$\arcsec$ in the sky plane.    
\end{itemize}

\section*{Discussion}
The individual components of repeating pairs resemble solar radio spikes \cite{Clarkson2023} in terms of time duration, frequency bandwidth, and image size, and likely share a similar generation mechanism. Spikes are suggested to be the result of non-thermal electron beams with smaller spatial sizes and lower densities than those that produce broadband emission such as type III bursts \cite{Melnik2014}, such that Langmuir wave growth and radio emission occur over narrow frequency ranges in a turbulent medium. 
Given that each component of the repeating pairs occurs at the same frequency, the likely explanation is that the bursts arise from the same emission source. Applying such a mechanism to the initial release of radiation then requires a secondary process to result in a D component. 

The presence of a D component shares similarities to drift pair bursts but are characteristically distinct due to drift rates that are 1--2 orders of magnitude lower, leading to total bandwidths that are an order of magnitude smaller and double the time separation between each burst. 
Compared to the E component, the D component is fainter, more diffuse, longer in decay time, and has lower drift rates, which are all signs of the radiation having undergone scattering. 

\begin{figure}[htbp]
\centering
\includegraphics[width=0.8\linewidth]{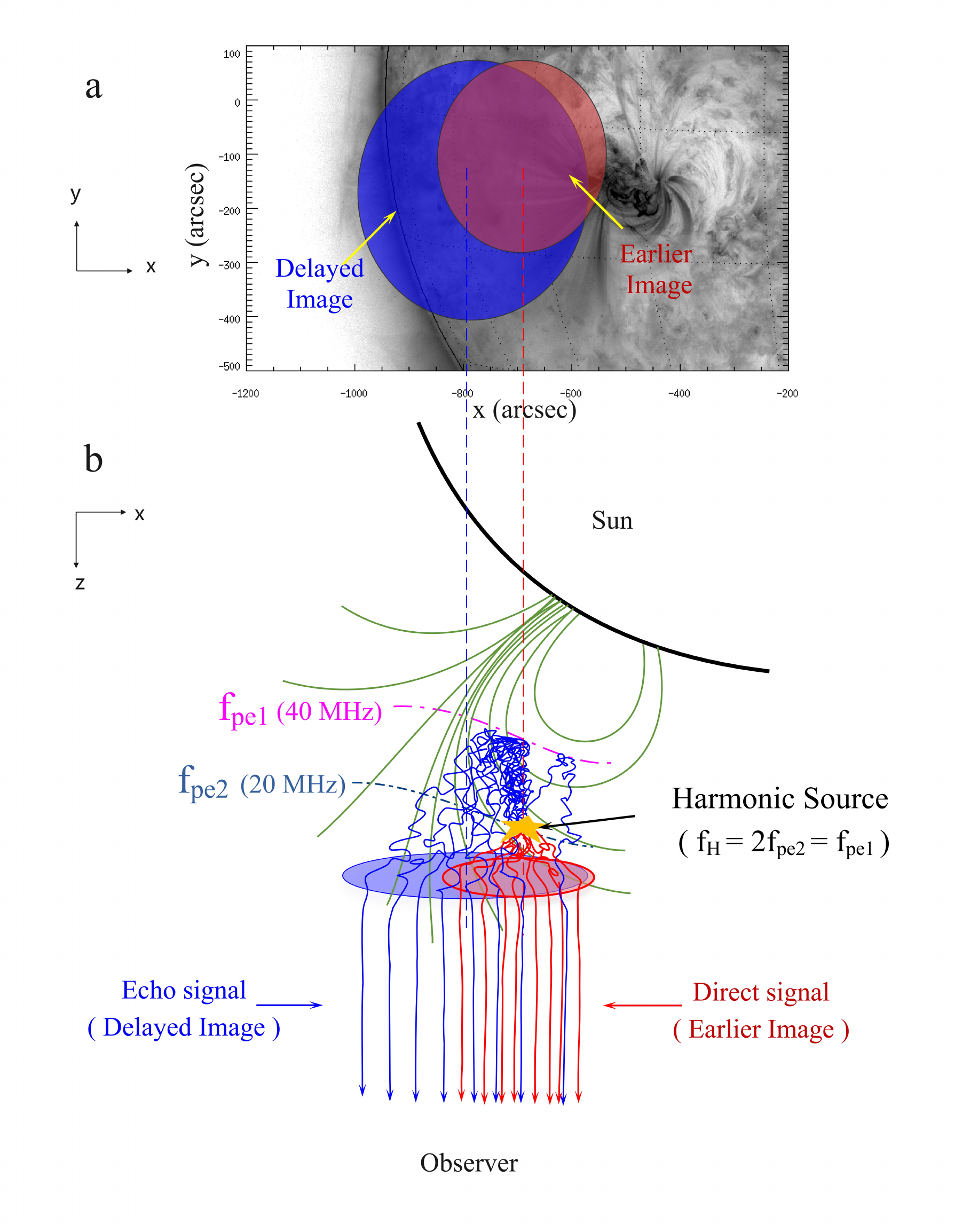}
\caption{\textbf{Potential mechanism for a repeating radio burst pair}. \textbf{a} The schematically shown earlier (E, red oval) and delayed (D, blue oval) images overlaid on the AIA image in the x-y plane. \textbf{b} A cartoon showing a potential mechanism for a repeating radio burst pair in x-z plane. The thick black curve represents the solar limb, and the green curves denote magnetic field lines. The gold star marks the actual radio source. The red and blue ovals indicate the apparent E and D sources, respectively, located at heights close to the $f_{\mathrm{pe2}}$ plasma layer. The red and blue curves trace the propagation paths of the radio emission. Radio emission at the harmonic frequency ($2f_{\mathrm{pe2}}$) from the actual source (gold star) propagates outward along the red trajectories, producing the direct scattered image that appears as the E component. Emission initially directed downward is reflected above the $f_{\mathrm{pe1}}$ layer (where $f_{\mathrm{pe1}} = 2f_{\mathrm{pe2}}$) and then propagates outward along the blue curves, giving rise to the D scattered image. }
\label{fig:cartoon}
\end{figure}

Therefore, we propose a possible mechanism as illustrated in Figure \ref{fig:cartoon}. 

In this scene, non-thermal electrons are generated through small-scale magnetic field reconnection, leading to the generation of Langmuir waves, which subsequently produce the radio source (indicated by a gold star in Figure \ref{fig:cartoon}) at the harmonic frequency ($2f_{pe2}$).
Radio emission propagating outward along the trajectories denoted by the red curves produces the direct scattered image. The radio emission initially directed downwards reflects backward above the plasma layer $f_{pe1}$ ($f_{pe1} = 2f_{pe2}$) and subsequently propagates outward along the trajectories marked by the blue curves and is responsible for the delayed scattered image. 
The difference in the propagation path produces a delay of about $4$~seconds, as evident from the simulations. This $4$~s delay  corresponds to the density scale height of \(R_\odot/4 - R_\odot/3 \), with larger scale heights yielding shorter delays.
The anisotropic scattering guides the radio waves predominantly along the magnetic field, whilst the increased time within the strong scattering region produces a more extended delayed source compared to the earlier source. 

\begin{figure}[htbp]
\centering
\includegraphics[width=1\linewidth]{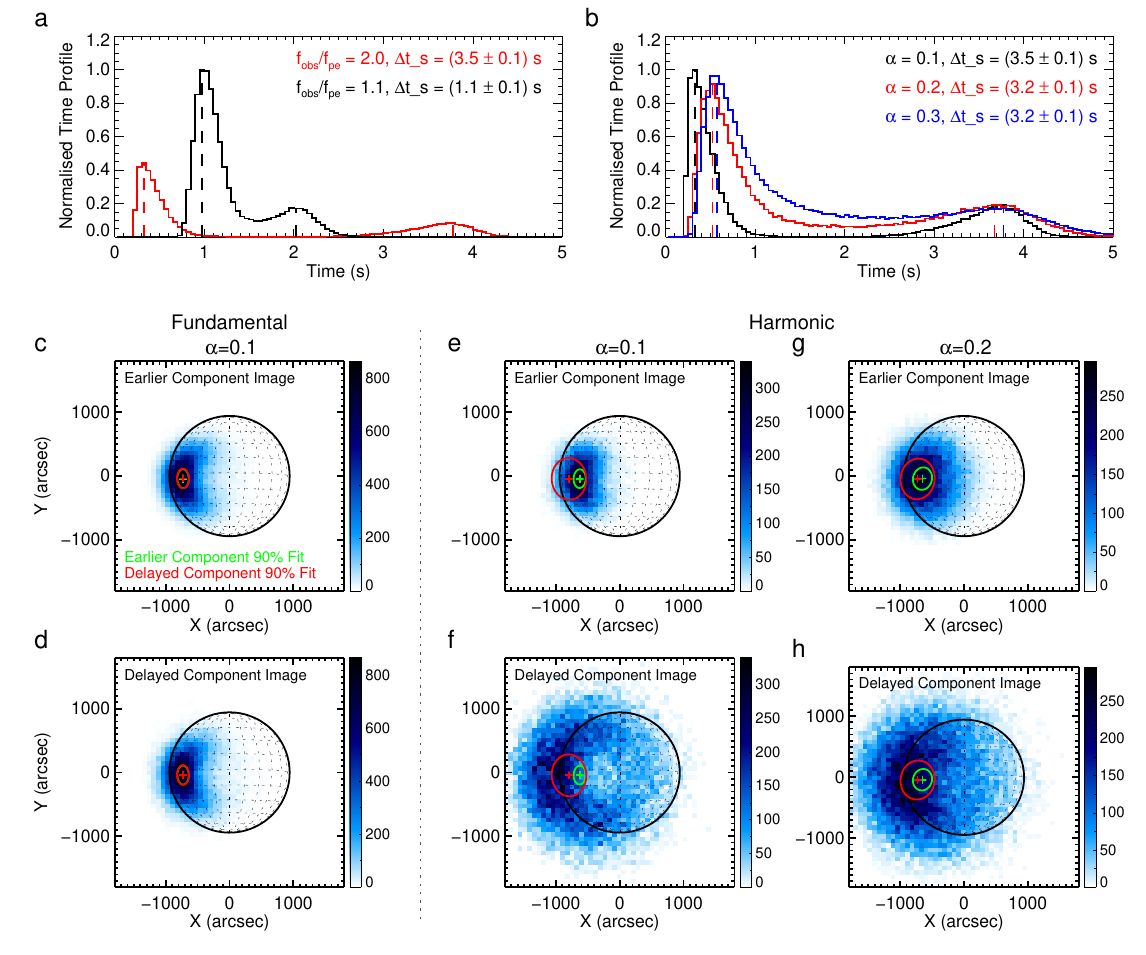}
\caption{\textbf{Simulated time profiles and images for 40 MHz emission}. \textbf{a} Comparison of time profiles for fundamental and harmonic sources with $\alpha=0.1$. \textbf{b} Comparison of harmonic sources for different anisotropy factors. The uncertainties in $\Delta t$ correspond to the standard error estimated from error propagation. \textbf{c-g} Simulated images of observed photons throughout the FWHM of the earlier (\textbf{c, e, g}) and delayed (\textbf{d, f, h}) components, represented as 2D histograms with bin sizes of 75$''$. The number of photons in each bin increases from light to dark blue, as shown by the colour bars. Panels (\textbf{c}, \textbf{d}) show a fundamental emission source with an anisotropy factor of 0.1, and panels (\textbf{e-h}) show harmonic emission sources with anisotropy factors of 0.1 (\textbf{e, f}) and 0.2 (\textbf{g, h}). In each image, the 90\% level of a fitted 2D Gaussian to both earlier and delayed component images is shown, with the crosses marking the fitted centroid locations.}
\label{fig:simul}
\end{figure}

It has been shown that strong anisotropy with an anisotropy parameter $\alpha\simeq[0.1-0.2]$ can produce highly directive emission \cite{Kontar2019} that can result in a turbulent radio echo \cite{Kuznetsov2020} due to reflection at the plasma frequency surface. Smaller values of the anisotropy parameter correspond to stronger directivity of the scattered radio waves.
Using the simulation code described by
\cite{Kontar2019, Kontar2023},  we perform radio-wave propagation simulations of 40 MHz fundamental (1.7~R$_\odot$) and harmonic (2.0~R$_\odot$) point sources injected at a heliocentric angle of $-15\degree$ from the disk center.
In each simulation, the anisotropy value is constant with height. The density profile follows a spherically-symmetric model suitable for active regions
as discussed in \cite{Kontar2023}. 
Figure \ref{fig:simul}a presents the time profiles for the fundamental and harmonic sources with a strong anisotropy factor of $\alpha=0.1$. 

In the case of fundamental emission, the time separation between the earlier and D components is 1~s, and the centroids of the two components are co-spatial as shown in Figure~\ref{fig:simul}c, inconsistent with the observed 4~s delay and imaging (Figure~\ref{fig:fig3}).
However, for the harmonic emission, the time separation is extended to about 4~s and the centroids are displaced between each component as observed. 
Considering harmonic emission with different values of $\alpha$  between 0.1--0.3 shows a weak effect on the time separation up to 3.5--3.2~s (Figure \ref{fig:simul}b), and a smaller displacement between the centroids for weaker anisotropy (Figure \ref{fig:simul}c). 
On this basis, we suggest that the mechanism producing the observed time separation, $4\times$ greater than that of drift-pair bursts \cite{Kuznetsov2020}, and the offset centroids between components can be explained by the repeating spike pairs being harmonic emission with strong anisotropy. The increased distance from the emitting location to the plasma-frequency surface where reflection occurs produces a greater separation between the E and D components and an extended apparent source for the D component.

The repeating pairs are associated with active region AR12665, which remained highly active throughout the observation interval, exhibiting both stronger core eruptions and continuous, weak mini-jets and outflows. While the temporal and spatial resolution of AIA (12~s cadence, $0.6\arcsec$ resolution) precludes a direct correlation with the sub-second bursts, insights from \citet{Chen2020} provide a compelling analogue. Their study demonstrated that single and clustered mini-jets launched from a tornado-like prominence arise from multi-site, small-scale magnetic reconnection occurring at high coronal altitudes. By inference, similar small-scale reconnection events could serve as the trigger for the repeating bursts, accelerating non-thermal electrons and driving plasma emission.

In several RPs, the E component displays a weaker amplitude than the D component. This contrast likely stems from variations in the propagation direction of the electromagnetic waves relative to Earth: a misaligned propagation vector would reduce the observed flux. Furthermore, significant deviation from the source-observer axis by either component would render the repeating pair undetectable, leaving only an isolated spike or stria. Our observations indeed reveal numerous individual spikes and spike chains, consistent with this scenario. This interpretation is reinforced by radio-wave simulations, which show that emission escapes preferentially along directions imposed by the magnetic field. We infer that repeating pairs demand more restrictive magnetic conditions for detection than spike bursts-an issue that awaits further investigation.

In this work, we have presented the first observations of solar spike-like RPs, a class of fine-scale radio structures revealed by LOFAR. More than one thousand such pairs were detected above active region AR12665 during a two-hour interval, demonstrating that these phenomena are both abundant and previously overlooked. 
Each pair consists of two short-lived, narrow-band bursts occurring at the same frequency:  an E component followed approximately four seconds later by a D component.
Our statistical analysis shows clear systematic differences between the two components. 
The delayed bursts are generally weaker, broader in duration, and exhibit markedly reduced frequency drift rates, while maintaining nearly identical spectral bandwidths and central frequencies to the earlier bursts. 
Imaging reveals that the E components originate closer to the active region core, whereas the D components are spatially displaced outward and spread over a larger area in the corona.

These combined temporal, spectral, and spatial characteristics point to an interpretation in which the D component represents a scattered echo of the earlier emission. The strong reduction in drift rate and the broader, fainter appearance of the delayed bursts are consistent with additional scattering along a downward and upward propagation path. 
Recent theoretical work on anisotropic scattering of fine structures supports such drift-rate reduction and enhanced path-dependent broadening. 
In this scenario, the bursts originate from harmonic emission at heights of roughly one solar radius above the solar surface, significantly above typical flare sites, 
implying that electron acceleration and magnetic reconnection occur at unexpectedly large coronal altitudes.
Importantly, the simulations also show that fundamental is inconsistent with the observed time delay and spatial displacement, clearly demonstrating that the emission is harmonic, which is a significant finding given that solar radio burst observations are often ambiguous about whether plasma emission is at the fundamental or harmonic.

The strong directivity of the emission further suggests that anisotropic coronal turbulence, aligned with the magnetic field, channels the scattered signal along preferred directions. The simulations provide strong support to previous works suggesting a density fluctuation anisotropy of 0.1-0.2, and their reflection time delay gives a measurement of density scale height. This has important implications for long-standing questions regarding radar echoes from the Sun: the same anisotropic scattering that guides the natural echo away from most viewing geometries may likewise suppress radar reflections.

Overall, these observations provide new evidence that small-scale reconnection and electron energization operate high in the corona, and that anisotropic scattering plays a key role in shaping the appearance of fine-structure radio bursts. The discovery of repeating pairs opens a new window into coronal plasma turbulence, magnetic geometry, and particle acceleration, and offers a promising diagnostic for future solar radio studies.

\section*{Methods}
\subsection*{Repeating Pair Screening and Spectral Parameter Measurements.}
A semi-automatic screening method was implemented to select repeating pair candidates and perform spectral parameter measurements. The following outlines the precise steps to be taken:
\begin{enumerate}
\item Initial Screening for repeating pairs. Candidate repeating pairs were initially identified based on their time-frequency occupancy. The analysis utilized raw LOFAR data to generate dynamic spectra with a duration of two minutes, a time cadence of 0.1~s, and a spectral resolution of 50~kHz. 
Each two-minute dynamic spectrum underwent manual inspection to detect isolated repeating pairs or extended chains of repeating pairs.
\item High-Resolution Dynamic Spectral Analysis. Following the identification of time-frequency coordinates for each burst pair, dynamic spectra were reconstructed using LOFAR data at full time resolution (10~ms) 
and frequency resolution (12.2~kHz) to enable detailed analysis.
\item Definition of Time-Frequency Boundaries and Profile Extraction. For each individual RP, time and frequency boundaries were defined. Each high-resolution spectrum (see example in Figure~\ref{fig:fig2}a) was displayed, and the integrated flux as a function of frequency was calculated over a manually selected time interval using a two-point click method. This facilitated the examination of both E and D bursts (highlighted by yellow lines in Figure~\ref{fig:fig2}a. Flux peaks and troughs were subsequently extracted, as illustrated in Figure~\ref{fig:fig2}b.
\item Measurement of Time and Frequency FWHM.
Fitting: To characterize the temporal profile shape of bursts, an asymmetrical Gaussian function, previously employed in \cite{Gerekos2024}, was adopted. 
This model accounts for a Gaussian rise followed by an exponential decay at a given peak frequency. Fitting was performed using the mpcurvefit routine with a five-parameter function:
\begin{equation}
    \begin{split}
        g(a,b,\mu,\sigma,\lambda;x) &= a\int_0^x\frac{1}{\sigma\sqrt{2\pi}}e^{-\frac{1}{2}(\frac{x-\bar{x}-\mu}{\sigma})^2}e^{-\lambda\bar{x}}\,\mathrm{d}\bar{x}+b\\
        &= a\frac{\lambda}{2}\exp\left\{{\frac{\lambda}{2}\left(2\mu+\lambda\sigma^2-2x\right)}\right\}\left(1-\mathrm{erf}\left\{\frac{\mu+\lambda\sigma^2-x}{\sqrt{2}\sigma}\right\}\right)+b 
    \end{split}
\end{equation}
where $a$ controls the overall scaling of the burst, $b$ controls the background level, $\mu$ is the mean of the Gaussian sector of the function, $\sigma^2$ is its variance, and $\lambda$ controls the decay rate of the exponential sector. The error function is defined as $\mathrm{erf}(z)\equiv 2\pi^{-1/2}\int_0^z e^{-t^2}\,\mathrm{d}t$. As exemplified in Figure \ref{fig:fig2}c and zoomed-in Figure \ref{fig:fig2}d, the rise time $\tau_r$ and decay time $\tau_d$ are then computed as the two HWHMs of the fitted curve. The peak time and the full width at half maximum (FWHM), given by the sum of the rise and decay times ($\tau_r + \tau_d$), can be determined for each burst.
Spectral Profile Fitting: The frequency profile at the peak time was fitted with a Gaussian function to determine the peak frequency and the corresponding FWHM (Figure \ref{fig:fig2}e).

\item Calculating the Frequency Drift Rate: 
The frequency drift rate was derived using the following procedure (see Supplementary Figure~1). From the dynamic spectrum, the frequency-integrated flux was computed and subsequently fitted with a Gaussian function. The times $t_1$ and $t_2$, corresponding to the FWHM of the Gaussian fit, defined the interval for calculating the frequency drift. For each time step $t_i$ within $[t_1, t_2]$, the flux density was plotted as a function of frequency. The peak frequency—i.e., the frequency at which the flux density attains its maximum—was extracted either directly from the frequency-flux profile or from its Gaussian fit, with the latter approach being particularly advantageous in the presence of frequency gaps. 
This procedure yielded a peak frequency value for each time step between $t_1$ and $t_2$. The resulting peak frequencies exhibited an approximately linear trend, allowing the drift rate to be estimated through linear regression. Although some bursts displayed a variable drift rate (see Supplementary Figure~2), a linear fit was applied to all bursts in this study to ensure consistency and facilitate comparison.
\end{enumerate}

\subsection*{Radio Source Position.}
\begin{figure}[htbp]
\centering
\includegraphics[width=0.9\linewidth]{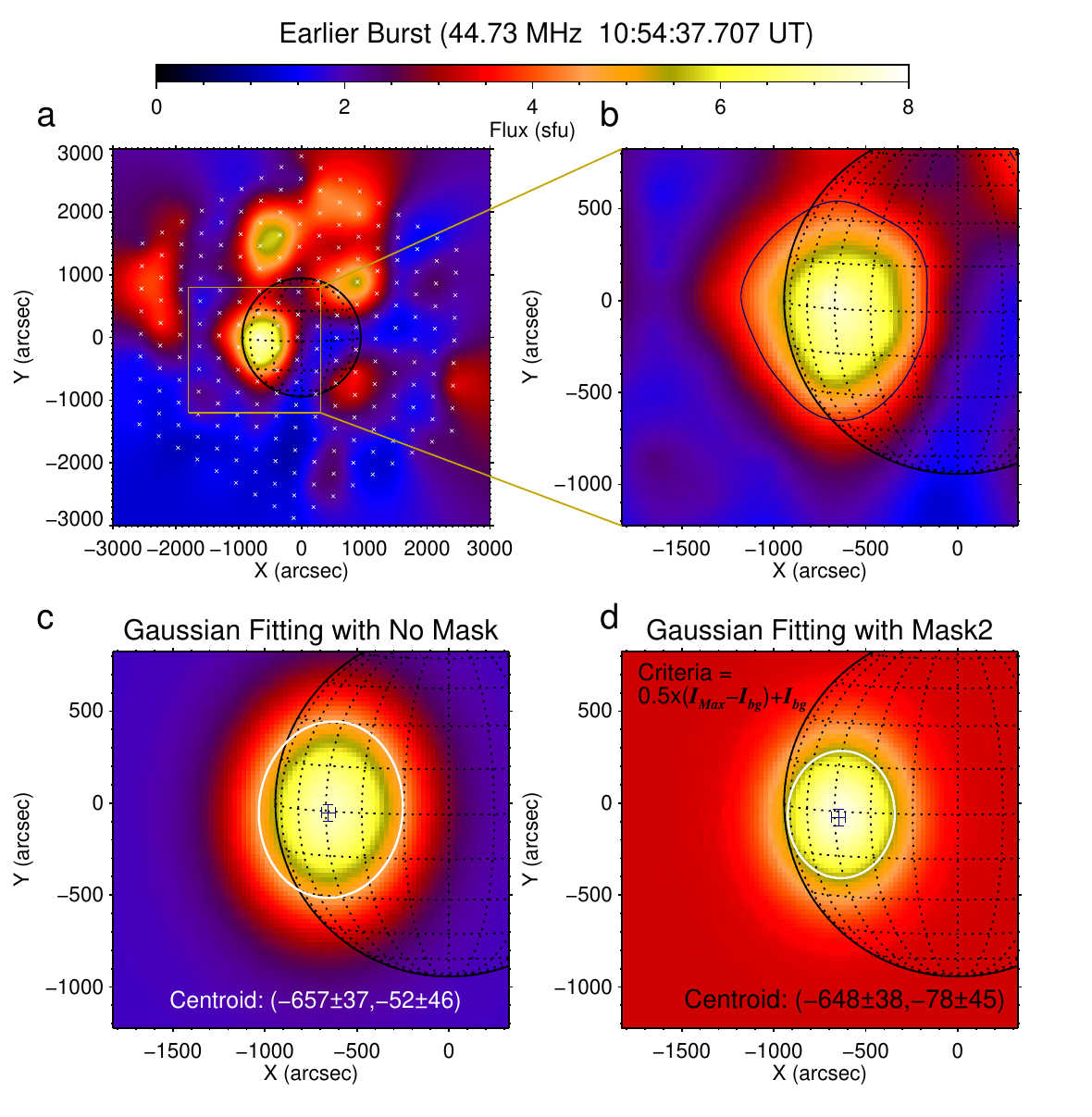}
\caption{\textbf{LOFAR image of the earlier burst at 10:45:38 UT and 44.73 MHz and 2D Gaussian fit results}. \textbf{a} Full disk LOFAR image. The white crosses show the phased array beam locations. The orange rectangle indicates the field of view of panel \textbf{b}. \textbf{b} Zoomed image used for Gaussian fitting, the black contour indicates the half maximum flux. \textbf{c} Restructured image based on the result from an elliptic Gaussian fitting without masking. \textbf{d} Restructured image based on the elliptic Gaussian fitting with a mask of $0.5\times(I_{Max} - I_{bg}) + I_{bg}$. The white oval in panels \textbf{c} and \textbf{d} indicates the FWHM level, and the plus symbol with error bars (respective 1$\sigma$ fitting error) marks the centroid of each burst.} 
\label{fig:fig10}
\end{figure}

The LOFAR images utilized in this work are based on observations 
from the 24-core Low Band Antenna stations with tied-array beam forming mod. 
The maximum baseline of the core station is around 3.5 km, 
which provides an angular resolution of about 390$\arcsec$ at 45 MHz. 
The array of 217 tied-array beams covers the sky out to 2$\rsun$ with a mosaic beam spacing of 0.1 degrees, as shown in Figures \ref{fig:fig10}a. 

As with the maps of Tau A \cite{Gordovskyy2019}, the maps of the Sun (Figure \ref{fig:fig10}a) also demonstrate typical patterns for interferometric observations, one bright source and several other fainter sources known as side lobes. As mentioned in \cite{Gordovskyy2019}, the position of the main lobe does not shift significantly with frequency at frequencies (30--50)~MHz. However, as the frequency increases, the fainter sidelobes become brighter. 
The frequency range of the repeating pairs presented in the work covers (38--60)~MHz. To partially eliminate the influence of the side lobes, the field-of-view (FOV) is restricted to a range of $-1800\arcsec$ to $300\arcsec$ along the x-axis and $-1200\arcsec$ to $800\arcsec$ along the y-axis, as indicated by the red window in Figures \ref{fig:fig10}a, which encompasses the primary lobe. 
The clipped LOFAR images, as shown in Figures \ref{fig:fig10}b, are used for 2D Gaussian fitting. 
To obtain the centroid position of bursts at frequencies 38-50 MHz, we adopt the fitting with the 2D tilted Gaussian function as used in \cite{Kontar2017}, \begin{equation}
S(x,y) = S_0\exp\left(-\frac{x^{\prime2}}{2\sigma_x^2}-\frac{y^{\prime2}}{2\sigma_y^2}\right),
\end{equation}
 where $x^{\prime} = (x-x_s)\cos(T)-(y-y_s)\sin(T)$ and $y^{\prime} = (x-x_s)\sin(T) + (y-y_s)\cos(T)$, where $T$ is the rotation from the $x$ axis in the clockwise direction. 
We determine the Gaussian parameters for each time and frequency bin by minimizing the chi‑squared statistic:
\begin{equation}
\chi^2 = \sum_{i=1}^N \frac{\bigl(F_i - S(x_i^b, y_i^b; S_0, x_s, y_s, \sigma_x, \sigma_y, T)\bigr)^2}{\delta F^2},
\end{equation}
where the sum runs over all \(N\) beam positions. The fitted parameters are: $S_0$ (peak amplitude), the source centroid coordinates $(x_s, y_s)$, the rms extents $\sigma_x$ and $\sigma_y$, and $T$, the rotation angle of the ellipse relative to the $x$-axis. For each frequency and beam, the background flux level before the burst is taken as the flux uncertainty $\delta F$ (typically about 1~sfu). The half‑maximum area of the Gaussian fit yields the source's areal extent. The positional errors $\delta x_s$ and $\delta y_s$ for the source location $(x_s, y_s)$ are given by \cite{Condon1997}:
\begin{equation}
\delta x_s \approx \sqrt{\frac{2}{\pi}} \frac{\sigma_x}{\sigma_y} \frac{\delta F}{S_0} h, \qquad 
\delta y_s \approx \sqrt{\frac{2}{\pi}} \frac{\sigma_y}{\sigma_x} \frac{\delta F}{S_0} h,
\end{equation}
with $h$ representing the angular resolution.

In the context of Gaussian fitting for radio bursts, a clear distinction 
exists between bursts with stronger and weaker peak fluxes. 
For bursts exhibiting a peak flux greater than 5 sfu, convergence 
in the Gaussian fitting process is typically attainable without the necessity for a masking technique, as evidenced in Figure \ref{fig:fig10}c. Conversely, for bursts with a weaker peak flux, specifically those below 3 sfu, the implementation of a mask becomes essential to ensure convergence of the fitting results.
In particular, for the delayed burst depicted in Supplementary Figure~3b, a masking criterion of 1.5 sfu serves as the minimum threshold required to achieve convergence in the Gaussian fitting process. 
The application of different masking values can significantly influence the fitting results, which can be identified in Figures \ref{fig:fig10}(c, d).
The plus symbols accompanied by error bars in these figures denote the centroids and their respective 1$\sigma$ error, while the white ovals indicate the FWHM level. Notably, an increase in the masking value correlates with a substantial decrease in the FWHM, whereas the centroid experiences only minor variations.
For instance, as depicted in Figures~\ref{fig:fig10}(c, d), the FWHM decreases from approximately $760\arcsec$ to $570\arcsec$ along the x-axis and from $940\arcsec$ to $680\arcsec$ along the y-axis. The centroid of the source transitions from approximately $(-657 \pm 37,\ -52 \pm 46)\arcsec$ without a mask to $(-648 \pm 38,\ -78 \pm 45)\arcsec$ when employing a mask defined as $0.5 \times (I_{\mathrm{Max}} - I_{\mathrm{bg}}) + I_{\mathrm{bg}}$. Similarly, Supplementary Figure~3(c, d) presents fitting results using a mask with a criterion of 1.5\,sfu (red contour) and a mask defined as $0.5 \times (I_{\mathrm{Max}} - I_{\mathrm{bg}}) + I_{\mathrm{bg}}$ (black contour). The centroids corresponding to these masks are approximately $(-891 \pm 99,\ -17 \pm 132)\arcsec$ and $(-896 \pm 102,\ -80 \pm 128)\arcsec$, respectively. Ideally, the fitted Gaussian centroid should remain consistent across varying mask criteria; however, the observed differences suggest that the source may not conform to an ideal Gaussian distribution. Thus, excluding the FWHM, the centroid derived using a mask defined as $0.5 \times (I_{\mathrm{Max}} - I_{\mathrm{bg}}) + I_{\mathrm{bg}}$ likely proves to be more reasonable and effective for weaker bursts. Consequently, the centroids presented in this paper are estimated through a 2D Gaussian fitting methodology, utilizing the aforementioned mask criteria.

\section*{Data Availability}
The source data required to reproduce all figures in this study are published alongside this paper. 
The raw observational LOFAR datasets generated and analyzed during the current study are publicly accessible from \url{https://www.astro.gla.ac.uk/users/eduard/L599637/}, which correspond to the LOFAR Long Term Archive under project code LC8\_027 (\url{https://lta.lofar.eu/Lofar?mode=projects_page&group=Cycle%208}). 
The SDO data were retrieved from \url{https://sdo.gsfc.nasa.gov/data/}, and the PFSS data were obtained from \url{http://www.lmsal.com} using the SolarSoftWare (SSW) package. 

\section*{Code Availability}
The IDL software for working with LOFAR data is available as part of the SolarSoftWare (SSW) package at \href{https://sohowww.nascom.nasa.gov/solarsoft/radio/lofar/}{https://sohowww.nascom.nasa.gov/solarsoft/radio/lofar/}. The simulation code used for radio wave transport in the solar wind is based on the \texttt{radio\_waves} package originally developed by Kontar et al.~\cite{Kontar2019,Kontar2023}. The exact version of the simulation code used in this paper has been permanently archived and is available as a fixed release~\cite{Ma2026radio_waves}. The original repository is available at \href{https://github.com/edkontar/radio_waves/}{https://github.com/edkontar/radio\_waves/} under the MIT license.

\bibliography{main.bib}

\section*{Acknowledgements}
This paper is based (in part) on data obtained with the LOFAR telescope (LOFAR-ERIC)  under project code LC8\_027. LOFAR~\cite{vanHaarlem2013} is the Low Frequency Array designed and constructed by ASTRON. It has observing, data processing, and data storage facilities in several countries that are owned by various parties (each with its own funding sources) and that are collectively operated by the LOFAR European Research Infrastructure Consortium (LOFAR-ERIC) under a joint scientific policy. The authors are also thankful to the SDO/AIA team and SDO/HMI team for the data, and to LMSAL for providing data and tools to compute, visualize, and analyze PFSS extrapolations. 

\section*{Funding}
S.M. discloses support for this research from the National Key R\&D Program of China (NRDPC) under grants 2022YFF0503800 and 2021YFA1600500. E.P.K. and D.L.C. acknowledge support via STFC/UKRI grant ST/Y001834/1. E.P.K. is supported by the Leverhulme Trust (Research Fellowship RF-2025-357). C.H. discloses support for this research from NRDPC grant 2021YFA1600500/2 and the Strategic Priority Research Program of the Chinese Academy of Sciences under grant XDB0560000.  Y.Y. acknowledges NRDPC grant 2021YFA1600500/3 and the Tianchi Talent Program of the Xinjiang Uygur Autonomous Region of China. S.M., C.H., and Y.Y. were also supported by the Specialized Research Fund for the State Key Laboratory of Solar Activity and Space Weather. This work benefited from an international team grant (\url{http://www.issibern.ch/teams/lofar/}) from ISSI Bern, Switzerland. 

\section*{Author contributions}
S.M. conceived and coordinated the study and performed the data analysis. E.P.K. contributed to the study design, provided the foundational code, and guided the result interpretation. D.L.C. developed the code, conducted the simulations, and aided in result interpretation. H.C. carried out the SDO data analysis and contributed to figure preparation and interpretation of results. Y.Y. contributed to the interpretation of the results. All authors participated in writing and revising the manuscript.

\section*{Ethics declarations}
\subsection*{Competing interests}
The authors declare no competing interests.

\section*{Supplementary material}

\begin{figure}[ht!]
\centering
\includegraphics[width=0.9\linewidth]{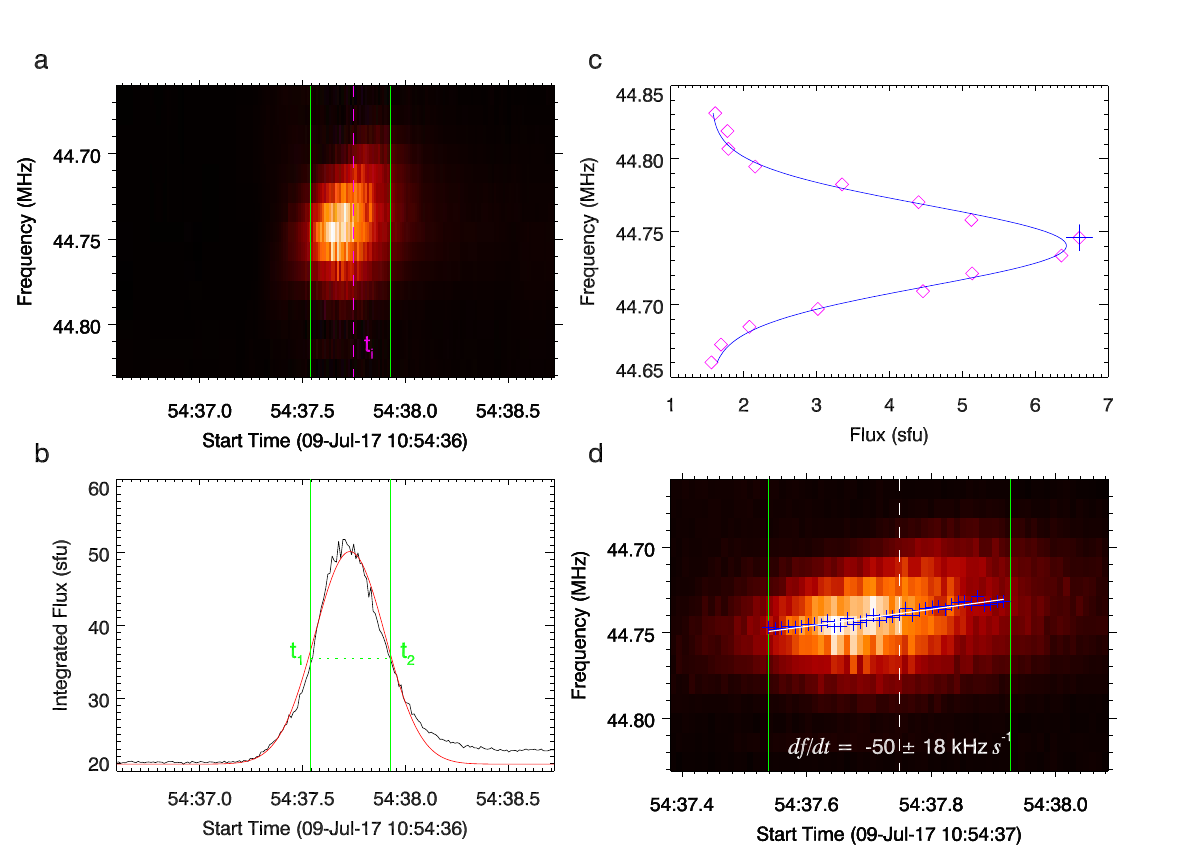}
\caption{\textbf{Supplementary Figure 1: Method to calculate the frequency drift rate of a radio burst.} \textbf{a} Initial dynamic spectrum, \textbf{b} Integrated flux over frequency, \textbf{c} Frequency flux profile at time $t_i$, \textbf{d} Dynamic spectrum with fitted peak frequencies at each time from $t_1$ to $t_2$ and linear fitting to obtain the frequency drift rate (the white line). The vertical green lines indicate the time borders where the integrated flux exceeds its half maximum. The diamonds in panel b mark the flux over frequency at time $t_i$, and the blue curve shows the Gaussian fitting. The blue plus symbols indicate the peak frequencies.}
\label{supfig:1}
\end{figure}

\begin{figure}[ht!]
\centering
\includegraphics[width=0.9\linewidth]{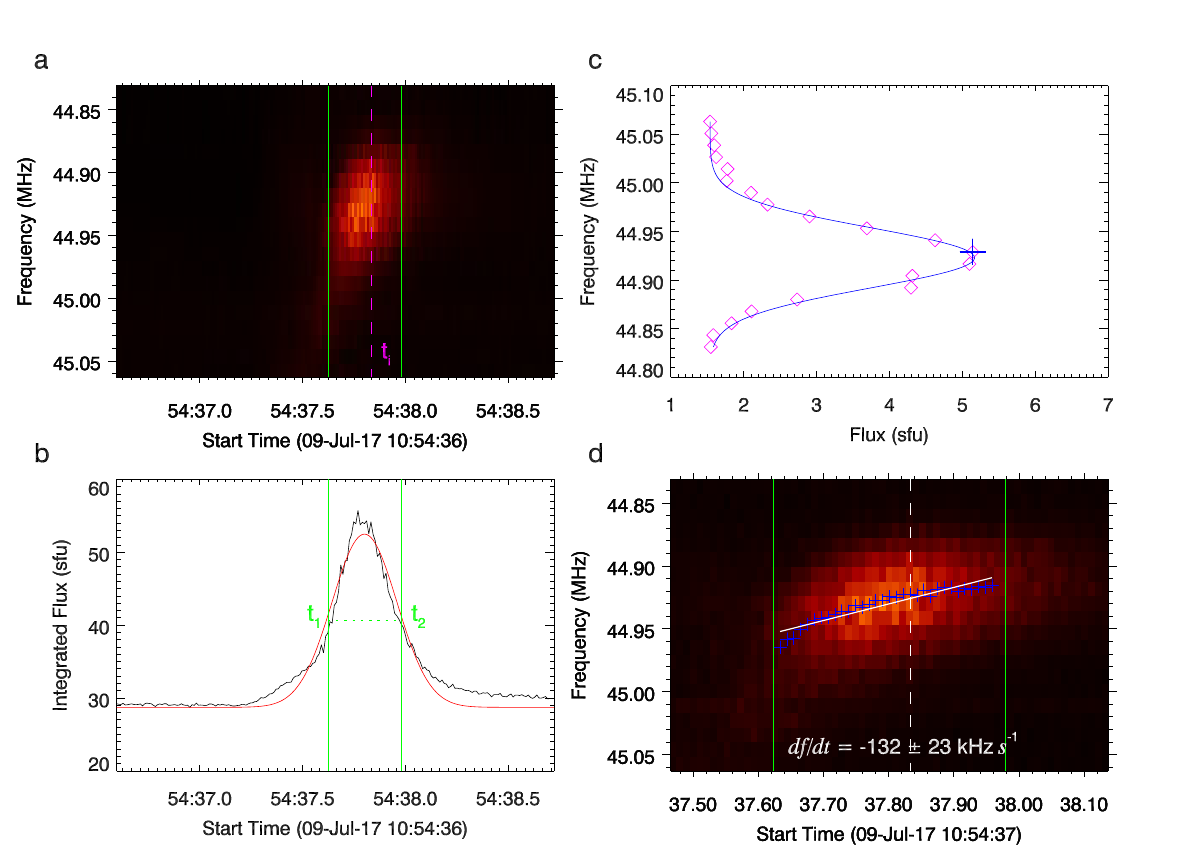}
\caption{\textbf{Supplementary Figure 2: Method to calculate the frequency drift rate of another radio burst.} \textbf{a} Initial dynamic spectrum, \textbf{b} Integrated flux over frequency, \textbf{c} Frequency flux profile at time $t_i$, \textbf{d} Dynamic spectrum with fitted peak frequencies at each time from $t_1$ to $t_2$ and linear fitting to obtain the frequency drift rate (the white line). The vertical green lines indicate the time borders where the integrated flux exceeds its half maximum. The diamonds in panel \textbf{b} mark the flux over frequency at time $t_i$, and the blue curve shows the Gaussian fitting. The blue plus symbols indicate the peak frequencies.}
\label{supfig:2}
\end{figure}

\begin{figure}[ht!]
\centering
\includegraphics[width=0.9\linewidth]{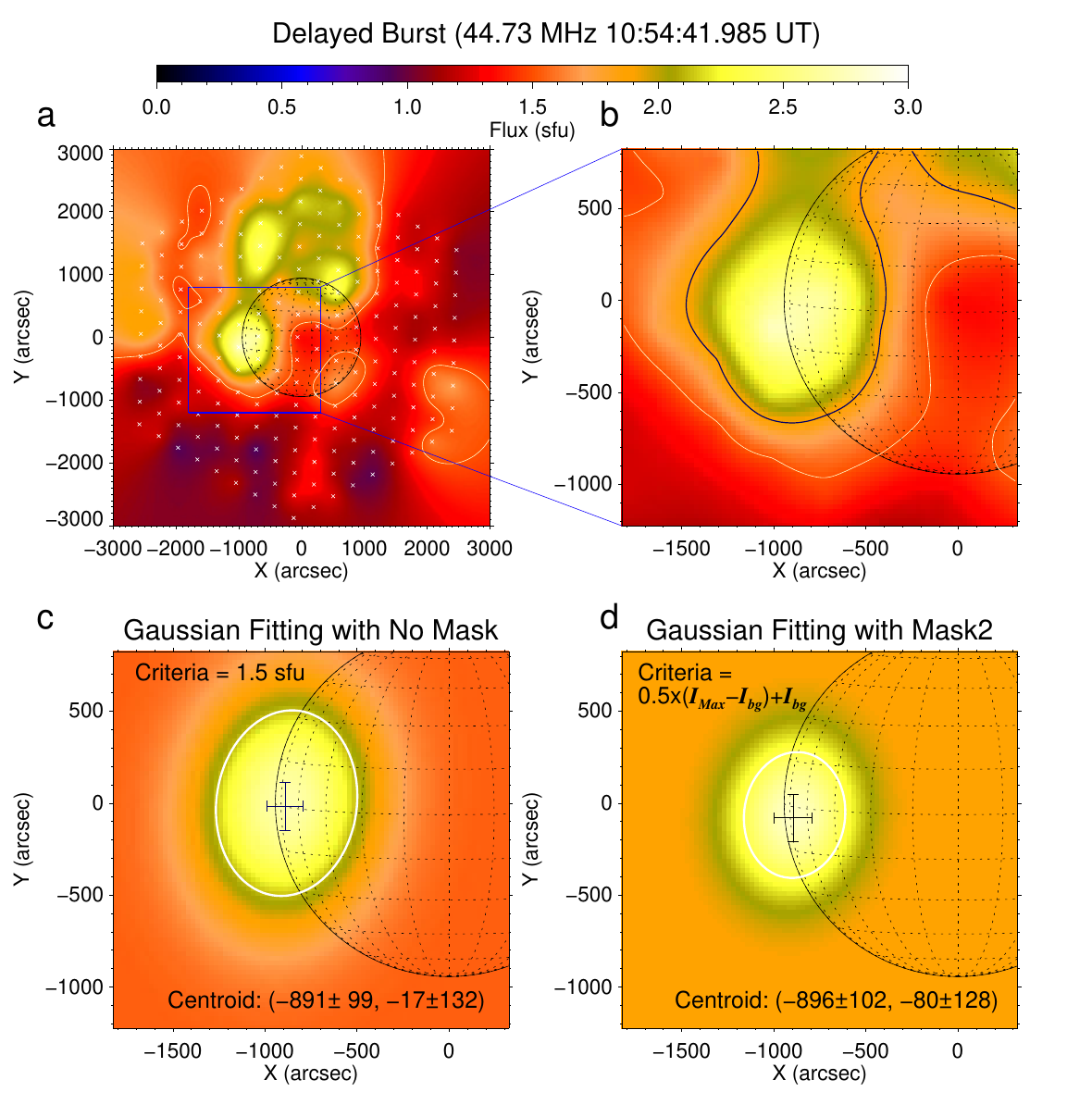}
\caption{\textbf{Supplementary Figure 3: LOFAR image of the delayed burst at 10:54:42 UT and 44.73 MHz and 2D Gaussian fit results.} \textbf{a} Full disk LOFAR image. The white crosses show the phased array beam locations. The blue rectangle indicates the FOV of panel b. \textbf{b} Zoomed image used for Gaussian fitting, the black contour indicates the half maximum flux. \textbf{c} Restructured image based on the result from an elliptic Gaussian fitting without a mask of 1.5 sfu. \textbf{d} Restructured image based on the elliptic Gaussian fitting with a mask of $0.5\times(I_{Max} - I_{bg}) + I_{bg}$. The white oval in panels \textbf{c} and \textbf{d} indicates the FWHM level, and the plus symbol with error bars marks the centroid of each burst.}
\label{supfig:3}
\end{figure}

\clearpage

\renewcommand{\arraystretch}{1.2}
\begin{longtable}{|c|l|c|S[table-format=3.0]|S[table-format=5.2]|S[table-format=4.2]|S[table-format=4.2]|S[table-format=4.2]|S[table-format=4.2]|}
\caption*{\textbf{Supplementary Table 1: Summary of statistical parameters used in box-and-whisker plots.}$^*$} \\
\hline
Panel & Quantity & Dataset & {N} & {5\%} & {25\%} & {50\%} & {75\%} & {95\%} \\
\hline
\endfirsthead

\caption*{\textbf{Supplementary Table 1 (continued)}} \\
\hline
Panel & Quantity & Dataset & {N} & {5\%} & {25\%} & {50\%} & {75\%} & {95\%} \\
\hline
\endhead

\hline
\multicolumn{9}{|r|}{Continued on next page} \\
\endfoot

\hline
\multicolumn{9}{|p{0.95\linewidth}|}{
\footnotesize
$^*$ \textbf{Definition of parameters.}
$\Delta t$ – Full Width at Half Maximum (FWHM) durations, where subscripts $E$ and $D$ denote earlier and delayed components (this rule applies to all following definitions). 
$\Delta t_s$ – time separation between the peak fluxes of the E and D components; 
$\Delta t_E/\Delta t_D$ – time duration ratio;
$\Delta f_E$, $\Delta f_D$ – FWHM bandwidths; $\Delta f_s$ – separation in frequency between the flux peaks of the E and D components;
$I_{0E}$, $I_{0D}$ – peak fluxes; 
$(df/dt)_E$, $(df/dt)_D$ – frequency drifts.
``Total'' – full bursts spanning 38–60~MHz; ``Image'' – bursts observed in 38–48~MHz that were successfully imaged with LOFAR.
} \\
\endlastfoot

Fig.4(a) & $\Delta t_E$ (s)               & Total   & 350 & 0.19 & 0.25 & 0.31 & 0.43 & 0.68 \\
         &                                 & Image   & 162 & 0.17 & 0.26 & 0.32 & 0.41 & 0.67 \\
         & $\Delta t_D$ (s)               & Total   & 304 & 0.62 & 0.82 & 1.00 & 1.20 & 1.45 \\
         &                                 & Image   & 158 & 0.71 & 0.94 & 1.12 & 1.30 & 1.58 \\
\hline
Fig.4(b) & $\Delta t_s$ (s)                & Total   & 241 & 3.47 & 4.03 & 4.23 & 4.39 & 4.52 \\
         &                                 & Image   & 121 & 3.97 & 4.13 & 4.28 & 4.41 & 4.51 \\
\hline
Fig.4(c) & $\Delta t_E/\Delta t_D$ & Total & 241 & 0.18 & 0.25 & 0.30 & 0.38 & 0.56 \\
         &                                 & Image   & 121 & 0.17 & 0.23 & 0.28 & 0.33 & 0.52 \\
\hline
Fig.4(d) & $\Delta f_E$ (kHz)              & Total   & 466 & 45.38 & 60.62 & 77.04 & 103.92 & 153.89 \\
         &                                 & Image   & 219 & 42.26 & 58.36 & 73.24 & 102.15 & 148.83 \\
\hline
Fig.4(e) & $\Delta f_D$ (kHz)              & Total   & 393 & 49.84 & 67.36 & 81.44 & 105.85 & 153.74 \\
         &                                 & Image   & 197 & 46.62 & 62.07 & 79.29 & 107.26 & 149.52 \\
\hline
Fig.4(f) & $\Delta f_s$ (kHz)              & Total   & 340 & -8.24 & 1.64 & 6.16 & 9.54 & 15.43 \\
         &                                 & Image   & 158 & -7.46 & 1.61 & 5.70 & 9.71 & 16.33 \\
\hline
Fig.4(g) & $I_{0E}$ (sfu)                 & Total   & 350 & 2.50 & 4.03 & 6.78 & 12.68 & 30.17 \\
         &                                 & Image   & 162 & 2.29 & 3.47 & 4.83 & 8.26 & 21.50 \\
\hline
Fig.4(h) & $I_{0D}$ (sfu)                 & Total   & 304 & 1.82 & 2.31 & 2.96 & 3.74 & 6.26 \\
         &                                 & Image   & 158 & 1.79 & 2.08 & 2.78 & 4.15 & 6.38 \\
\hline
Fig.4(i) & $I_{0D}/I_{0E}$   & Total & 341 & 0.10 & 0.26 & 0.44 & 0.76 & 1.43 \\
         &                                 & Image   & 174 & 0.12 & 0.34 & 0.54 & 0.96 & 1.59 \\
\hline
Fig.5(c) & $(df/dt)_E$ (kHz/s)            & Total   & 404 & -170.8 & -68.2 & -31.6 & 4.8 & 72.2 \\
         &                                 & Image   & 188 & -119.2 & -66.8 & -33.8 & 0.3 & 41.9 \\
\hline
Fig.5(d) & $(df/dt)_D$ (kHz/s)            & Total   & 300 & -21.7 & -13.0 & -7.8 & -3.5 & 3.8 \\
         &                                 & Image   & 146 & -22.5 & -11.4 & -6.5 & -3.1 & 4.4 \\
\end{longtable}

\section*{Source data}
Source Data.
Supplementary movie. 

\end{document}